\documentclass[prd,showpacs,showkeys,superscriptaddress,amsmath,preprintnumbers,nofootinbib]{revtex4}

\usepackage{array}          
\usepackage{graphicx}       
\usepackage{dcolumn}        
\usepackage{bm}             
\usepackage{amsfonts}       

\def \be{\begin{equation}}
\def \ee{\end{equation}}
\def \bea{\begin{eqnarray}}
\def \eea{\end{eqnarray}}
\def \met{\mbox{g}}
\begin{document}

\title{A Toy Model for Testing Finite Element Methods to Simulate
Extreme-Mass-Ratio Binary Systems}

\author{Carlos F. Sopuerta}
\affiliation{Institute for Gravitational Physics and Geometry,
Penn State University, University Park, PA 16802}
\affiliation{Center for Gravitational Wave Physics,
Penn State University, University Park, PA 16802}
\affiliation{Department of Astronomy \& Astrophysics,
Penn State University, University Park, PA 16802}
\author{Pengtao Sun}
\affiliation{Department of Mathematics,
Penn State University, University Park, PA 16802}
\affiliation{Center for Computational Mathematics and Applications,
Penn State University, University Park, PA 16802}
\affiliation{Department of Mathematics,
Simon Fraser University, Burnaby B.C. V5A 1S6, Canada}
\author{Pablo Laguna}
\affiliation{Institute for Gravitational Physics and Geometry,
Penn State University, University Park, PA 16802}
\affiliation{Center for Gravitational Wave Physics,
Penn State University, University Park, PA 16802}
\affiliation{Department of Astronomy \& Astrophysics,
Penn State University, University Park, PA 16802}
\affiliation{Department of Physics, Penn State University,
University Park, PA 16802}
\author{Jinchao Xu}
\affiliation{Institute for Gravitational Physics and Geometry,
Penn State University, University Park, PA 16802}
\affiliation{Department of Mathematics,
Penn State University, University Park, PA 16802}
\affiliation{Center for Computational Mathematics and Applications,
Penn State University, University Park, PA 16802}

\date{\today}

\begin{abstract}
Extreme mass ratio binary systems, binaries involving stellar mass objects
orbiting massive black holes, are considered to be a primary source of
gravitational radiation to be detected by the space-based interferometer
LISA.  The numerical modelling of
these binary systems is extremely challenging because the scales
involved expand over several orders of magnitude. One needs to
handle large wavelength scales comparable to the size of the massive
black hole and, at the same time, to resolve the scales in the
vicinity of the small companion where radiation reaction
effects play a crucial role.   Adaptive finite element methods,
in which quantitative control of errors is achieved automatically
by finite element mesh adaptivity based on posteriori error
estimation, are a natural choice that has great potential for achieving the
high level of adaptivity required in these simulations.
To demonstrate this, we present the results of simulations of a toy model,
consisting of a point-like source orbiting a black hole under the action of a
scalar gravitational field.
\end{abstract}



\preprint{CGWP/xx}

\maketitle

\section{Introduction}
As we enter the era of Gravitational Wave Astronomy, a number of experiments
to detect and study gravitational waves have been set up and some
others are presently under development (see~\cite{Hough:2005gw} for a recent
account).   Among the second group we find
one of the experiments that is presently attracting a considerable amount of
attention: the Laser Interferometer Space Antenna
(LISA)~\cite{Vitale:2002qv,Danzmann:2003ad,Danzmann:2003tv,Prince:2003aa},
a collaboration between ESA and NASA that is scheduled to be launched in the
next decade. Extreme-Mass-Ratio Binaries
(EMRBs) are considered to be a primary source of gravitational radiation to be
detected by LISA~\cite{Barack:2003fp,Gair:2004ea}.  They consist of a
``small'' object, such a main sequence star, a stellar mass black hole, or a
neutron star, with mass $m$ ranging from $1 M_\odot$ to $10^2 M_\odot$, orbiting
a massive black hole (MBH) with mass $M$ ranging from $10^3 M_\odot$ (if we
consider the case of intermediate mass black holes) to $10^9 M_\odot$
(the case of big supermassive black holes sitting in the center of galaxies).
This translates to EMRBs with mass ratios, $\mu = m/M\,,$ in the range
$10^{-3} - 10^{-9}\,$.
In order to exploit this type of systems through LISA, it is crucial to have a
good theoretical understanding of the evolution of these systems, good enough
to produce accurate waveform templates in support of data analysis efforts.

Because there is no significant coupling between the strong curvature effects
between the MBH and its companion, relativistic perturbation theory is a well
suited tool to study EMBRs.  Clearly, the accuracy of this approximation
depends on the smallness of the ratio $\mu$.
The goal is to study the perturbations generated by the small body in the
(background) gravitational field of the MBH, and how these perturbations affect
the motion of the small body itself, which follows the geodesics of the
perturbed spacetime.   That is, one is after studying how the presence of the
small body affects its own trajectory.  This problem is usually known in
literature as the {\em radiation reaction problem}.  This is an old problem and
several approaches to deal with it have been proposed (see the recent reviews
by Poisson~\cite{Poisson:2004lr,Poisson:2005rc}, Detweiler~\cite{Detweiler:2005sd}
and Mino~\cite{Mino:2005yw,Mino:2005zb,Mino:2005uf}).  A pragmatic approach
is to use energy-momentum balance arguments~\cite{Tanaka:1993pu,Cutler:1994pb,
Hughes:1999bq,Hughes:2000ss,Hughes:2001jr,Glampedakis:2002cb}.
Under this approach, one estimates the changes in the small body {\em constants
of motion} by computing the fluxes of energy and angular momentum at infinity
and through the MBH horizon.
This approach works well in the adiabatic regime, when the time scale of the
radiation reaction is much bigger than the orbital time scale.  Until now it
has dealt with special orbits of the Kerr black hole, but it has not yet
produced results for generic orbits because of the difficulty of adjusting
the third constant of motion, the Carter constant, present in generic
geodesics of the Kerr spacetime. However, there have been some recent advances
in this direction~\cite{Mino:2003yg,Hughes:2005qb,Drasco:2005is,Sago:2005gd}.

An alternative approach consists in trying to describe the radiation reaction
effects on the small body as the action of a local {\em self-force} that is
responsible for the deviations from the geodesic motion.  A rigorous formulation
of this concept has been given by the first time by Mino, Sasaki and
Tanaka~\cite{Mino:1997nk}, and later, adopting an axiomatic approach, by Quinn
and Wald~\cite{Quinn:1997am}.  These works give a formal
prescription to compute the self-force.  For the practical implementation of
these prescription some techniques have been proposed (see~\cite{Lousto:2005co}
for a recent progress report): the {\em mode-sum}
scheme~\cite{Barack:1999wf,Barack:2000eh,Barack:2001bw,Barack:2001gx}, and a
regularization scheme based on zeta-function regularization
techniques~\cite{Lousto:1999za}.

In this paper we want to advocate an alternative, and at the same time complementary,
approach: The direct numerical integration of the (linearized) evolution equations
for the gravitational field together with the evolution equations for the small body,
which form a system of partial differential equations (PDEs) coupled to
a system of ordinary differential equations (ODEs).  We can already find in the
literature an attempt to use numerical methods
for simulating EMRBs~\cite{Bishop:2003bs}.  In this work, the {\em perturbations}
are computed by using the full general relativistic equations in the framework
of the characteristic formulation (in contrast to a Cauchy-type formulation),
with the small body being described by an energy-momentum distribution
that moves ``rigidly'' along a geodesic of the numerically computed
spacetime.  The main drawback of this work is the size of the small body, which
is bigger than the MBH horizon.  Another approach that has been used in the
literature~\cite{Lopez-Aleman:2003ik} is to describe the gravitational field
by using the Teukolsky formalism~\cite{Teukolsky:1973st} implemented in the time
domain (using a numerical code introduced in~\cite{Krivan:1997hc})
and by modelling the small body by smearing the singularities in the source
term by the use of narrow Gaussian distributions.
This work has also the problem that the size of the small body is too big
(in comparison with the size of the MBH).  This shows the main underlying
difficulty in the numerical simulation of EMRBs, namely, the problem
involves a vast range of physical scales (spatial and temporal) that expand over
several orders of magnitude. Specifically, one needs to handle not only large
wavelength scales comparable to the massive black hole, but also to resolve the
scales in the vicinity of the small object where radiation reaction effects
play a crucial role.

An obvious conclusion we can extract from these facts is that, in order to
carry out successful numerical simulations of EMRBs, we need a high
degree of adaptivity.  Our proposal is the use of the Finite Element Method (FEM)
as a natural choice to achieve this high level of
adaptivity.
Finite Element Methods have not been used occasionally in General Relativity
(see~\cite{Arnold:1998pr,Metzger:2004pr})
To demonstrate that these numerical techniques have great potential of
leading to successful simulations of EMRBs, we present results from simulations
of a toy problem consisting of a point-like source orbiting a black hole in
scalar gravitation in $2+1$ dimensions.   Our aim is to test FEM techniques
in a simple representative problem that possesses the main ingredients and
challenges of the astrophysical EMRBs.  That is, at this stage we are
basically conducting a feasibility study.
In our toy model, the spacetime metric is fixed and aims at describing the
gravitational field of a non-rotating black hole.  For computational efficiency,
we have made a reduction from three spatial dimensions to two.  In this
reduction, which we explain in detail later, the metric we work with is not
longer a solution of the Einstein vacuum equations but it keeps the important
property that its geodesics coincide with the equatorial geodesics of the
Schwarzschild spacetime.
This metric is not dynamical but fixed.   The dynamical gravitational field
is described by a scalar field on this spacetime, satisfying a wave-like
equation with a source term that describes the presence of the small object.
An important ingredient of our model is the use of a particle description
for the small object.
The equations of motion of this particle are the geodesics of the fixed
spacetime metric modified by the presence of (spatial) components of the
gradient of the gravitational scalar field.
In this way we have that the particle orbits the black hole
subject to radiation reaction: indeed, the particle generates the scalar
gravitational field which affects its own motion.

In this work we compare numerical simulations that use the simple classical
FEM with simulations that use an adaptive-mesh FEM.
The essence of the adaptive mesh technique is to produce
real-time local mesh coarsening or refinement to achieve the desired
level of smoothness in the solution.  To that end, a good posteriori error estimator to
predict the regions in the computational domain where rapidly changes take
place is extremely important.  There are several ways to approach this problem.
Theoretical research on this subject~\cite{Chen_Sun_Xu_2003} has found that
the Hessian matrix of the numerical solution can accurately predict where the
steepest gradients of the solution would take place.   We demonstrate that this
technique when applied to our toy model easily captures the dynamics of
the field in the vicinity of the particle since around the particle location
either the field or the source term will change much more than anywhere
else.  We then refine the local area surrounding the particle and resolve it
more accurately so that we are able to achieve our final target.

The plan of this paper is the following: In Section~\ref{sec2} we
introduce the particular theoretical model we want to study, namely: the description
of the gravitational field and of the point-like source, the computational
setup, and an energy balance test that can be used to test the numerical computations.
In Section~\ref{sec3} we describe the computational techniques that we use
for the time-domain simulations of our theoretical model:
the FEM, the discretization of our equations using Finite
Elements, the numerical method for solving the equations of motion of the particle,
and finally, the Adaptive Finite Element Method (AFEM).   In Section~\ref{sec4} we present
and discuss the numerical results of the simulations.  Here, we distinguish between
the simulations that use the classical FEM and those that use the AFEM.
We devote the last Section~\ref{conclusions} to discuss
the main results of this work and the future perspectives that it opens.

\section{Theoretical aspects of the toy model\label{sec2}}
In the following, we describe in detail the main aspects of our toy model, namely:
(i) the gravitational theory we use, (ii) the derivation of the equations to be solved,
(iii) the particle's description, and (iv) the computational setup.
We use physical units in which $c=G=1\,.$ We use lowercase Greek letters for
spacetime indices and lowercase Latin letters for {\em spatial} indices.
In subsections~\ref{scagrav} and~\ref{decompo}
indices run as: $\mu\,,\nu\,,\ldots = 0-3\,;$ $i\,,j\,,\ldots = 1-3\,.$

\subsection{Scalar Gravity\label{scagrav}}
Scalar gravitation is a theory of gravity that although it cannot
we applied to our physical world (we know it cannot fit all the available
experimental data), is a good laboratory for numerical relativity due to its
simplicity.   Of particular interest for our purposes is the fact that
(scalar) gravitational waves exist in scalar gravity.
In this theory, the gravitational field is described by a scalar field
$\Phi(x^\mu)$ on a spacetime geometry described by a non-dynamical metric
tensor $\met_{\mu\nu}$, which we just prescribe
(the version presented in problem 7.1 of the textbook by
Misner, Thorne and Wheeler~\cite{Misner:1973cw} and
in~\cite{Shapiro:1993vr,Shapiro:1994xg,Scheel:1994xh} considers only the
case in which the non-dynamical spacetime is the flat spacetime).
The scalar field does not affect to the spacetime structure defined by
$\met_{\mu\nu}$.  We are interested in studying the evolution of a particle-like
object orbiting a black hole in this theory.  To that end, we consider a background
spacetime metric $\met_{\mu\nu}$ describing the geometry of a non-rotating
black hole, and a particle of mass $m$ that follows the worldline $z^\mu(\tau)$,
where $\tau$ denotes the particle proper time.  The action of this particle-field
system is
\begin{equation}
S = \int \sqrt{-g}{\cal L}d^4x =
-\int \sqrt{-g}\left(\frac{1}{8\pi}\met^{\mu\nu}\nabla_\mu\Phi
\nabla_\nu\Phi-\rho\mbox{e}^\Phi \right)d^4x \,, \label{scalar}
\end{equation}
where the comoving density $\rho$, describing the particle, is given by
\begin{equation}
\rho = \int\frac{m}{\sqrt{-g}}\delta^4[x^\mu-z^\mu(\tau)]d\tau
     = \frac{m}{u^t\sqrt{-g}}\delta^3[x^i - z^i(t)] \,,
\end{equation}
where $t$ is a time coordinate and $u^t$ is the corresponding component of
the velocity of the particle:
\begin{equation}
u^\mu = \frac{dz^\mu(\tau)}{d\tau}~\Longrightarrow~u^t = \frac{d t}
{d\tau}\,. \label{parvel}
\end{equation}
Varying the action~(\ref{scalar}) with respect to the scalar field $\Phi\,,$
we obtain the gravitational field equation:
\begin{equation}
\met^{\mu\nu}\nabla_\mu\nabla_\nu\Phi = 4\pi\mbox{e}^\Phi \rho \,.
\label{fe}
\end{equation}
Varying now the action with respect to $z^\mu\,,$ we obtain the particle's
equations of motion
\begin{equation}
u^\nu\nabla_\nu u^\mu +(\met^{\mu\nu}+u^\mu u^\nu)\nabla_\nu\Phi = 0\,.
\label{pe}
\end{equation}
Equations (\ref{fe}) and (\ref{pe}) form a coupled system of partial and ordinary
differential equations respectively.  Equation (\ref{fe}) is a hyperbolic
and nonlinear equation that describes the dynamical gravitational field whereas
(\ref{pe}) is a system of equations of motion for the particle.
In the absence of $\Phi$ the particle follows the geodesics of the
background spacetime $\met_{\mu\nu}$, but in the present of $\Phi$ it will
not longer follow the geodesic of the background.  The way in which
$\Phi$ influences the motion of the particle is through its gradient
projected orthogonally to the four-velocity of the particle.  At the
same time, the particle influences the gravitational field $\Phi$ through
the source term $\rho$.  The coupling between the particle and the field
is where the radiation reaction is encoded:  The particle induces a
nonzero $\Phi\,,$ and this field induces a deviation in the motion of the particle
from the background geodesic motion, and at the same this motion affects the
evolution of the field, and so on.   Therefore, the situation is the
same as in the general relativistic case where the particle is treated as a
perturbation on the background spacetime of the MBH, and where the radiation
reaction mechanism works in the same way.

This system has a well-defined total energy-momentum tensor $T_{\mu\nu}\,,$
which is given by
\begin{equation}
T^{\mu\nu} =
\frac{2}{\sqrt{-g}}\frac{\delta(\sqrt{-g}{\cal L})}{\delta \met_{\mu\nu}}\,.
\end{equation}
It has two differentiated components, one associated with the scalar
gravitational field, ${}^{(\Phi)}T_{\mu\nu}$, and the other with
the particle, ${}^{(\rho)}T_{\mu\nu}$: $T_{\mu\nu}=
{}^{(\Phi)}T_{\mu\nu}+ {}^{(\rho)}T_{\mu\nu}$.  Their expressions are
respectively
\begin{equation}
{}^{(\Phi)}T_{\mu\nu} = \frac{1}{4\pi}\left(\nabla_\mu\Phi\nabla_\nu\Phi
-\frac{1}{2}\met_{\mu\nu}\nabla^\sigma\Phi\nabla_\sigma\Phi \right) \,,
\end{equation}
\begin{equation}
{}^{(\rho)}T_{\mu\nu} = \rho\mbox{e}^\Phi u_\mu u_\nu \,.
\end{equation}
One can show that the energy-momentum conservation equation
\begin{equation}
\nabla_\mu T^{\mu\nu}=0\,, \label{energy}
\end{equation}
follows provided matter conservation holds, that is,
$\nabla_\mu(\rho u^\mu)=0$.

\subsection{3+1 decomposition of the equations\label{decompo}}
In order to numerically solve the equations (\ref{fe},\ref{pe}) it is
convenient to rewrite them in a 3+1 language, which allows for an initial-value
problem formulation.  To that end, we follow the usual $3+1$ decomposition
used in numerical relativity.  That is, we write our spacetime line-element as
follows
\begin{equation}
ds^2 = -\alpha^2dt^2+h_{ij}(dx^i+\beta^idt)(dx^j+\beta^jdt) \,. \label{2m1}
\end{equation}
where $h_{ij}$ is the spatial metric of the $\{t = const.\}$ hypersurfaces,
$\alpha$ denotes the lapse function and $\beta^i$ the shift vector.
The normal to the hypersurfaces is:
\begin{equation}
n_\mu = (-\alpha,0)\,,~~~n^\mu = \frac{1}{\alpha}(1,-\beta^i)\,.
\end{equation}
The covariant and contravariant components of the
spacetime metric are
\begin{equation}
(\met_{\mu\nu}) = \left(
     \begin{array}{c|c} -\alpha^2 + h_{ij}\beta^i\beta^j & h_{ij}\beta^j \\
                        \hline \\
                        \beta^ih_{ij} & h_{ij}
     \end{array} \right)\,,~~~~~
(\met^{\mu\nu}) = \left(
     \begin{array}{c|c} -\alpha^{-2} & \alpha^{-2}\beta^i \\
                        \hline \\
                        \alpha^{-2}\beta^j & h^{ij}-\alpha^{-2}\beta^i\beta^j
     \end{array} \right)\,, \label{stmetric}
\end{equation}
where $h^{ij}$ is the inverse of $h_{ij}$.  Then,
$h^{\mu}{}_{\nu} = \met^\mu{}_{\nu} + n^\mu n_\nu$
is the projector orthogonal to the hypersurfaces $\{t=const.\}$, which
also contains the induced metric that these hypersurfaces inherit:  Its spatial
components coincide with $h_{ij}$.

Equation~(\ref{fe}) for $\Phi$ is second-order in space and in time.
We are going to split it into two equations that are first-order in time.
To that end, we introduce the following definition:
\begin{equation}
\Pi = n^\mu\partial_\mu\Phi =
\frac{1}{\alpha}\left(\partial_t -\beta^i\partial_i \right)\Phi\,,
\end{equation}
Then, equation (\ref{fe}) can be split into the following two equations:
\begin{equation}
\left(\partial_t -\beta^i\partial_i \right)\Phi = \alpha\Pi\,, \label{fe1}
\end{equation}
\begin{equation}
\left(\partial_t -\beta^i\partial_i \right)\Pi = \alpha\left(
K \Pi + a^iD_i\Phi + \triangle\Phi -4\pi \mbox{e}^\Phi\rho\right) \,.
\label{fe2}
\end{equation}
In the second equation, $K$ denotes the trace of the extrinsic curvature
$K_{\mu\nu}$ of the hypersurfaces $\{t=const.\}$
\begin{equation}
K_{\mu\nu}= -\frac{1}{2}{\pounds}_n h_{\mu\nu}~~
\Longrightarrow ~~
K = -\nabla_\mu n^\mu \,. \label{k}
\end{equation}
The symbols $D_i$  and $\triangle$ denote the covariant derivative and
Laplacian associated with the induced metric $h_{ij}$. Then we can write
\begin{equation}
\triangle\Phi = h^{ij}D_iD_j\Phi =
\frac{1}{\sqrt{h}}\partial_i\left(\sqrt{h}\,h^{ij}\partial_j
\Phi \right) \,. \label{Laplacian}
\end{equation}
Moreover, $a^i$ denotes the spatial components of
$h^\mu{}_\nu n^\sigma\nabla_\sigma n^\nu$.  After some calculations
we find the following expression for $a^i$
\begin{equation}
a^i = h^{ij}D_j \ln\alpha = h^{ij}\partial_j \ln\alpha \,. \label{ai}
\end{equation}

Now, let us perform the 3+1 decomposition of the equations for the trajectory
of the particle.
Our coordinate system $(t,x^i)$ is, in principle, not adapted to the particle
trajectory in the sense that $t$ is not the proper time and $x^i$ are not
comoving coordinates.   Then, in this coordinate system the trajectory of the
particle is given by $(t,x^i=z^i(t))$, and the unit tangent velocity vector is:
\begin{equation}
u^\mu \partial_\mu = u^t \partial_t + u^i\partial_i \,.
\end{equation}
But also
\begin{equation}
u^\mu\partial_\mu = \partial_\tau~~~
\Longrightarrow~~~u^t = \frac{dt}{d \tau} \,,~~ u^i = \frac{dx^i}{d\tau}\,.
\end{equation}
Then, on the trajectory of the particle the following holds
\begin{equation}
u^i = \frac{dz^i(t)}{dt}\frac{d t}{d\tau} = v^i u^t\,,~~~
\mbox{where}~v^i = \frac{dz^i}{dt}\,.
\end{equation}
Due to the fact that $u^\mu$ is a unit timelike vector field
($\met_{\mu\nu}u^\mu u^\nu = -1$) we do not need to solve for all the
components.  One possibility is to solve the particle equations for the
quantities $(z^i(t),v^i(t))$.  Then, the equations that we get are:
\begin{eqnarray}
\frac{d}{dt}z^i &=& v^i\,, \label{peq1}  \\
\frac{d}{dt}v^i &=& f^i_{\mbox{\footnotesize g}} + f^i_\Phi \,, \label{peq2}
\end{eqnarray}
where
\begin{eqnarray}
f^i_{\mbox{\footnotesize g}} & = &
\frac{1}{(u^t)^2}\left(v^i\Gamma^t_{\rho\sigma}-\Gamma^i_{\rho\sigma}\right)
u^\rho u^\sigma \,,   \label{ag} \\
f^i_\Phi &=& \frac{1}{(u^t)^2}\left[v^i(\met^{t\rho}+u^tu^\rho)-
(\met^{i\rho}+u^iu^\rho)\right]
\partial_\rho\Phi \,, \label{af}
\end{eqnarray}
and the velocity component $u^t$ can be written in terms of $v^i$ and the
spacetime metric as follows
\begin{equation}
(u^t)^2 = -\left(\met_{tt}+2\met_{ti}v^i+\met_{ij}v^iv^j\right)^{-1} \,.
\end{equation}
The term $f^i_{\mbox{\footnotesize g}}$ gives the contribution to the geodesic
motion in the background spacetime $\met_{\mu\nu}$, whereas the term
$f^i_\Phi$ describe the deviation from the geodesic motion due to the action
of the scalar gravitational field $\Phi$.

\subsection{Pseudo-Schwarzschild Black Hole Background}
We want the background metric $\met_{\mu\nu}$ to represent the geometry of a
non-rotating black hole.  In 3+1 we can use the Schwarzschild metric.
However, to reduce the computational cost of our simulations, we reduce the
space of our toy model to two spatial dimensions. There are several ways in
which this reduction can be performed.  For instance, one can start directly
with the $2+1$ metric obtained from the Schwarzschild metric on the
hypersurface $z=0$.
Another possibility would be to start from the equations in the $3+1$ setup,
expanding all the different terms and then, to neglect all the derivatives with
respect to the coordinate $z$ and the components of the different objects in that
direction, and finally to set $z=0$.
In performing these dimensional reductions the metrics we obtain are no
longer solutions of Einstein's equations in the dimensionally-reduced spacetime,
but this is not an issue in our toy model since the spacetime metric is not
a dynamical object.  The important thing is that the metric we obtain from
these reductions keeps most of the important properties of the Schwarzschild
metric, in particular, the fact that from the two different reductions mentioned
above the equations for the geodesics are the same,
and they coincide with the geodesics of Schwarzschild in the plane $z=0$.
The equations for $\Phi$, when we introduce the expression of the
dimensionally-reduced metrics, would be in general different.
In this work we use the first possibility for the dimensional reduction.
As a consequence of this reduction the indices we use in this subsection
run as follows: $\mu\,,\nu\,,\ldots = 0-2\,;$ $i\,,j\,,\ldots = 1,2\,.$

In what follows we specify the form of the 2+1 background.
We start from the Schwarzschild metric in Cartesian Kerr-Schild coordinates
but reducing a space dimension.   That is, our background metric is described
by the following line element:
\begin{equation}
ds^2 = \eta_{\mu\nu}dx^\mu dx^\nu + 2H\ell_\mu\ell_\nu dx^\mu dx^\nu\,,
\label{ksmetric}
\end{equation}
where $\eta_{\mu\nu}=\mbox{diag}(-1,1,1)$ is the 2+1 Minkowski metric,
$\ell^\mu$ is a future-directed light-like vector field (both with respect to
the Minkowski and Schwarzschild metric), and $H$ is a scalar.  In Cartesian
Kerr-Schild coordinates these three objects are given by
\begin{equation}
\eta_{\mu\nu}dx^\mu dx^\nu = -dt^2 + \delta_{ij}dx^idx^j\,,~~~
\ell_\mu dx^\mu = -dt+ \frac{x_i}{r}dx^i\,,~~~
H = \frac{M}{r} \,,\label{kssch}
\end{equation}
where $M$  is the black hole mass and $r=\sqrt{\delta_{ij}x^ix^j}$.

We choose the $\{t=const. \}$ foliation of the 2+1 spacetime.  Then, the
values of the relevant 2+1 quantities that result from this choice are:
\begin{equation}
\alpha^2 = \frac{1}{1+2H}= \frac{1}{1+\frac{2M}{r}}\,, \label{met1}
\end{equation}
\begin{equation}
\beta^i = -\frac{2H}{1+2H}l^i\,,~~~~l^i = \frac{x^i}{r}\,, \label{met2}
\end{equation}
\begin{eqnarray}
h_{ij} = \delta_{ij} + 2Hl_il_j\,,~~~~~
h^{ij} = \delta^{ij} - \frac{2H}{1+2H}l^il^j\,, \label{met3}
\end{eqnarray}
\begin{equation}
\sqrt{h} = \sqrt{\det(h_{ij})} = \sqrt{1+2H} =
\sqrt{1+\frac{2M}{r}}\,,
\label{met4}
\end{equation}
\begin{equation}
K = -\frac{2M}{r^3}\left( 1+\frac{2M}{r}\right)^{-3/2} \,, \label{met5}
\end{equation}
\begin{equation}
a^i = -h^{ij}\frac{\partial_j H}{1+2H} = \frac{H}{r}\alpha^4 l^i =
\frac{1}{(1+\frac{2M}{r})^2} \frac{Mx^i}{r^3}\,. \label{met6}
\end{equation}
As it happens in any Kerr-Schild metric, the determinant of the spacetime metric
is equal to $-1\,,$ and hence we have the following relation between the lapse and
the determinant of the spatial metric: $\sqrt{h}= 1/\alpha\,,$ which agrees
with equations (\ref{met1},\ref{met4}).

The Dirac delta function that appears in the source $\rho$ is regularized by
using a Gaussian distribution:
\begin{equation}
\label{eq:delta}
\delta^2[x^i-z^i(t)]  \approx  \frac{1}{\left(\sqrt{2\pi}\,\sigma\right)^2}
\; e^{-\frac{R^2}{2\,\sigma^2}}\,,
\end{equation}
where
\begin{equation}
R^2 = \sum^2_{i=1}[x^i-z^i(t)]^2\,.
\end{equation}
The source function $\rho$ takes the form
\begin{equation}
\rho = \frac{m}{\left(\sqrt{2\pi}\,\sigma\right)^2\,u^t\,\alpha\sqrt{h}}
\; e^{-\frac{R^2}{2\,\sigma^2}} =
\frac{m}{\left(\sqrt{2\pi}\,\sigma\right)^2\,u^t}\;
e^{-\frac{R^2}{2\,\sigma^2}}\,. \label{source}
\end{equation}
As we have already mentioned, $m$ is the total rest mass of the particle.
We can recover this quantity, which is a constant of motion, from the
following expression:
\begin{equation}
m = \int\,\gamma\,\rho\,\sqrt{h}\,d^2x\,. \label{mass}
\end{equation}
where $\gamma=\sqrt{1+h^{ij}u_iu_j} = \alpha u^t\,.$
Note that by making the approximation (\ref{eq:delta}), the exact character
of this relation does not change due to the fact that the integral of the
Dirac delta and a given Gaussian over the whole two-dimensional domain yields
the same result.  On the other hand, the matter conservation relation
$\nabla_\mu(\rho\,u^\mu) = 0$ ensures that $m$, as defined by equation~(\ref{mass}),
is a conserved quantity.

The next thing we need is the explicit
form of the equations of motions for the particle,  which means that we need
to compute the right-hand side of equation~(\ref{peq2}), or equivalently,
the terms (\ref{ag},\ref{af}), for our spacetime metric~(\ref{ksmetric}).
The expressions that we obtain are:
\begin{eqnarray}
f^i_{\mbox{\footnotesize g}} &=& \frac{2M}{r^2}v^i\left\{-\frac{M}{r}+
\left(1+\frac{2M}{r}\right)\frac{x_j v^j}{r}+ \delta_{jk}v^jv^k-
\left(2+\frac{M}{r}\right)\frac{(x_j v^j)^2}{r^2}  \right\} \nonumber \\
& & - \frac{M}{r^3}x^i\left\{1-\frac{2M}{r}+\frac{4M}{r^2}x_j v^j +
2\delta_{jk}v^jv^k-\left(3+\frac{2M}{r} \right)\frac{(x_j v^j)^2}{r^2}
\right\}\,, \label{forceg}
\end{eqnarray}
\begin{eqnarray}
f^i_\Phi &=& \left[1-\delta_{jk}v^jv^k-\frac{2M}{r}\left(1-\frac{x_j v^j}{r}
\right)^2 \right]\left\{  \left[\frac{2M}{r^2}x^i-\left(1+\frac{2M}{r}\right)
v^i\right]\frac{\Pi}{\sqrt{1+\frac{2M}{r}}} \right. \nonumber\\
& & \left. +\frac{\frac{2M}{r}}{1+\frac{2M}{r}} \frac{x^i}{r}
\frac{x^j}{r}\partial_j\Phi - \partial^i\Phi \right\} \,, \label{forcep}
\end{eqnarray}
where we have used that
\begin{equation}
\left(u^t\right)^2 = \left[1-v^iv_i-\frac{2M}{r}\left(1-
\frac{x_iv^i}{r}\right)^2\right]^{-1} \,. \label{ut2}
\end{equation}
It is important to remark that the expression (\ref{forceg}) has the
same form as in the 3+1 case, confirming the fact that the geodesics
of our 2+1 spacetime coincide with the equatorial geodesics ($z=0$)
of the Schwarzschild spacetime.

\subsection{Toy Model Setup\label{toysu}}
To summarize, we have to solve a set of six equations.  Two of
them are PDEs, equations (\ref{fe1},\ref{fe2}), and contain a
non linear term, the one that introduces the coupling
between the gravitational scalar field and the matter sources.  The
other four equations are ODEs, equations~(\ref{peq1},\ref{peq2}),
describing the trajectory followed by the
particle of mass $m$.  The two sets of equations are coupled.

\begin{figure}[!htb]
\centerline{
{\includegraphics*[height=2.8in,width=3.in]{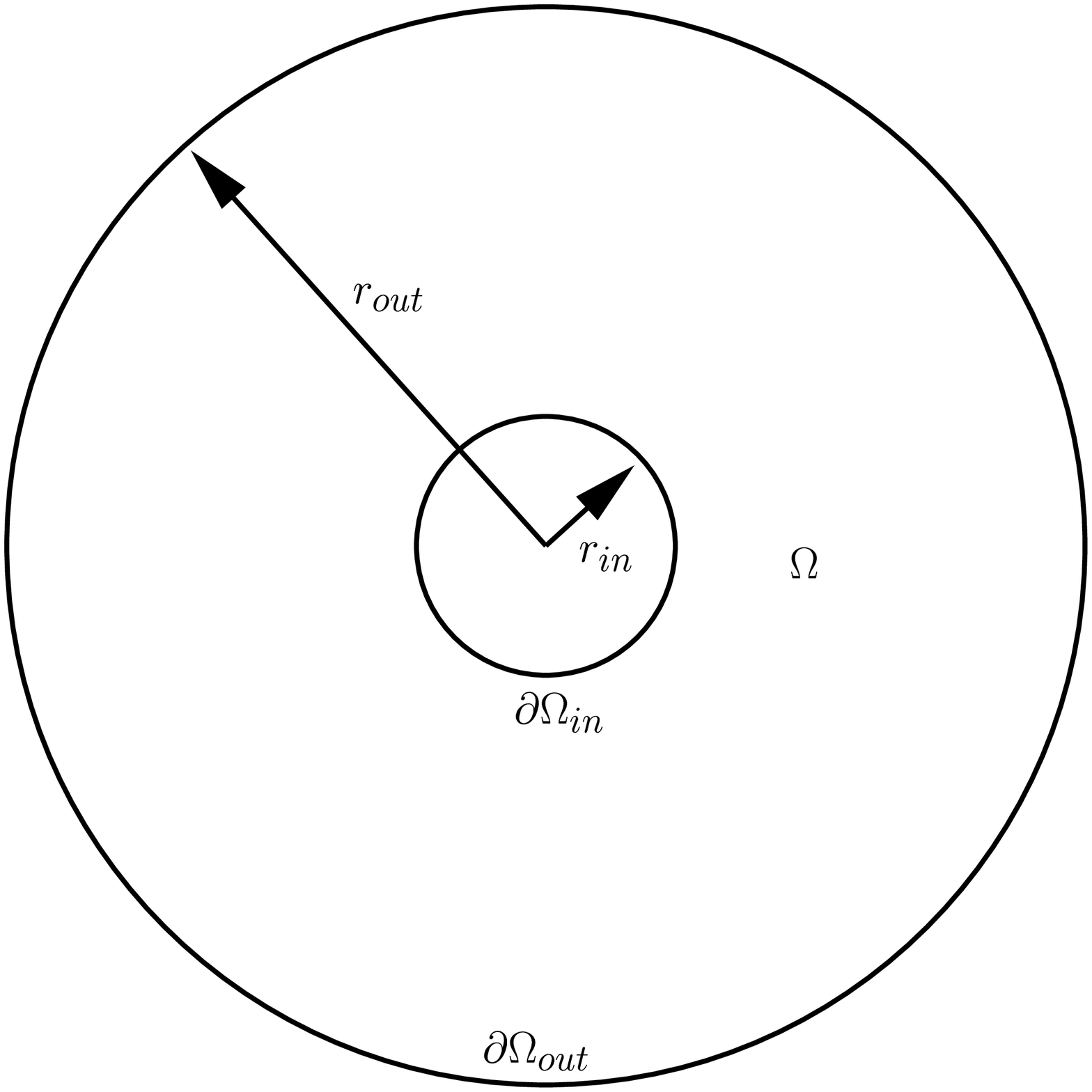}} }
\caption{Computational domain: $\Omega = \left\{(x,y)~|~ r_{in} <
r < r_{out} \right\}$.\label{domain}}
\end{figure}

To integrate these equations we are going to consider the following type of
spacetime domains: $\left[t_o,t_f\right]\times \Omega$, where $t_o$ and $t_f$
are the initial and final integration times, and $\Omega$ is the spatial domain
(see Fig.~\ref{domain}), the domain at every $\{t=const.\}$ slice,
consisting of two circular boundaries: an outer boundary
$\partial\Omega_{out}$ at $r=r_{out}>50M$ and a inner boundary
$\partial\Omega_{in}$ at $r=r_{in}<r_h$ where $r_h= 2M$  is the horizon
{\em radius}.  The aim is to locate $r_{out}$ far enough close to the
radiation zone so we can impose standard outgoing radiation boundary conditions.
With regard to the inner boundary, the idea is to {\em excise} the black hole
singularity from the computational domain, as it is done in many full numerical
relativity calculations of black hole dynamics~\cite{Shoemaker:2003td,
Sperhake:2003fc}, without affecting the computation of the field.  This can
be done as long as we have $r_{in}<r_h$ since the characteristics of the
PDEs (\ref{fe1},\ref{fe2}), for $r<r_h\,,$ all point in the direction of
$r=0\,,$ which entitles us to perform the
singularity excision.  Moreover, because of this property of the characteristics,
we do not need to impose any kind of boundary conditions at the inner boundary.
On the outer boundary
($r=r_{out}$) we use the two-dimensional Sommerfeld boundary
condition (see, e.g.~\cite{Bayliss:1980bt,Bayliss:1982bg}):
\begin{equation}
\left.  \left(\partial_t+\frac{x^i}{r}\partial_i+\frac{1}{2r}\right)
\Phi\right|_{r=r_{out}}= 0 \,. \label{bdc}
\end{equation}
We can rewrite this condition, using polar coordinates, like $(\partial_t+\partial_r)
(\sqrt{r}\,\Phi)|_{r=r_{out}}=0$.  However, in contrast to what happens in the
three-dimensional case, where $F(t-r)/r$ is an exact solution coinciding with the
radiative behaviour of the field (at large $r$), in the two-dimensional case
$F(t-r)/\sqrt{r}$ is not a solution of the equations, and therefore the boundary
condition (\ref{bdc}) does not capture correctly the radiative behaviour of the model.
In this sense, this boundary condition could be improved along the lines shown in
references~\cite{Bayliss:1980bt,Bayliss:1982bg}, by considering higher-order
derivative boundary conditions, or along the
works~\cite{Alpert:2000bt,Alpert:2002bt,Lau:2004jn,Lau:2004as}
where {\em exact} radiative boundary conditions are explored.

The initial data consists of two functions on $\mathbb{R}^2$,
$(\Phi_o,\partial_t\Phi_o)$ or equivalently $(\Phi_o,\Pi_o)$,
the initial position of the particle $z^i_o=z^i(t_o)$ and its
initial velocity $v^i_o=v^i(t_o)$.   For the scalar gravitational field we
are going to use the simplest initial data, which consists in setting
the initial value of $\Phi$ and its time derivative equal to zero
[$\Phi_o(x,y)\equiv\Phi(t=t_o,x,y)\,,\,
\dot\Phi_o(x,y)\equiv\partial_t\Phi(t,x,y)|_{t=t_o}$]:
\begin{equation}
\Phi_o(x,y)=0\,,~~~~~\mbox{and}~~~~~\dot\Phi_o(x,y)=0\,.
\end{equation}
In terms of $(\Phi,\Pi)$ this translates to
\begin{equation}
\Phi_o(x,y)=0\,,~~~~~\mbox{and}~~~~~
\Pi_o(x,y)=-\frac{\beta^i}{\alpha}\partial_i\Phi_o(x,y)\,.
\end{equation}
From a physical point of view, this data corresponds to a situation in which
the particle comes into
existence at the initial time $t=t_o$ and through the source term in equation
(\ref{fe2}) induces a non-zero scalar gravitational field.  A consequence of
this initial setup is the triggering of an spurious burst of radiation.

The particle's initial data, $(z^i_o,v^i_o)$,  is chosen
in such a way that it coincides with initial data that, in the absence of scalar
gravitational field, would correspond to circular geodesics of our background
spacetime~(\ref{ksmetric}).   Without loss of generality, we assume that
the particle is located initially at
\begin{equation}
z^i_o = (x_o,0)\,,
\end{equation}
where here we can prescribe freely the initial position $x_o$.
Then, the initial velocity is
\begin{equation}
v^i_o = \left(0,\sqrt{\frac{M}{x_o}}\right)\,.
\end{equation}
By using our knowledge of the geodesics [we recall that the geodesics
of~(\ref{ksmetric}) coincide with the equatorial geodesics of the 3+1
Schwarzschild spacetime], we can have an estimation of the time that the
particle takes for completing a circular orbit,
which we call ${\cal T}$.   This time can be obtained by using Kepler's third
law:
\begin{equation}
{\cal T}^2 = \frac{4\pi^2}{M}R^3 \,,
\end{equation}
where $R$ is the radius of the circular orbit.  It is important to remark
that this expression is exact when we neglect the effect of the scalar
gravitational field $\Phi$, otherwise it just gives an estimation for the
orbital period.
For an orbital radius $r=10\,M$ we have ${\cal T} \approx 198.7\,M\,.$

\subsection{Energy-balance Test}
An important feature of the toy model is the existence of a local
conservation law~(\ref{energy}) that involves both the energy-momentum
of the particle and of the dynamical scalar gravitational field.
This local conservation law together with
the symmetries of the spacetime lead to global conservation laws.  We
can use these global conservation laws as a test for the numerical
simulations of our toy model.

Let us derive the conservation law associated with the timelike Killing 
vector field that makes our spacetime static, ${\bm \xi}=\partial_t$.
Contracting equation (\ref{energy}) with $\xi^\mu$ and using the Killing 
equations, $\nabla_\mu\xi_\nu+\nabla_\nu\xi_\mu=0\,,$ we obtain:
\begin{equation}
\xi_\mu\nabla_\nu T^{\mu\nu} = 0~~\Rightarrow~~
\nabla_\nu\left(T^{\mu\nu}\xi_\mu\right) - T^{\mu\nu}\nabla_\nu
\xi_\mu = 0 ~~\Rightarrow~~ \nabla_\nu\left(T^{\mu\nu}\xi_\mu\right)
= 0\,.
\end{equation}
We are now going to integrate this relation over a compact region
of the spacetime, ${\cal V}\,,$ and we use the Gauss theorem to convert
volume integrals into surface integrals:
\begin{equation}
\int_{\cal V} \nabla_\nu\left(\xi_\mu T^{\mu\nu}\right)\,
{\bm d{\cal V}} = 0~~~~\Longrightarrow~~~~
\int_{\partial {\cal V}}T^{\mu\nu}\xi_\mu d\Sigma_\nu = 0\,,
\label{elaw}
\end{equation}
where $\partial {\cal V}$ is the boundary of ${\cal V}$, which is a
closed hypersurface,  and $d\Sigma_\mu$ the volume 1-form associated with 
$\partial{\cal V}$.
In this way we have obtained an energy-momentum conservation law that
tell us that the flux of the vector $T^\mu{}_\nu\xi^\nu$ across the boundary
of the region ${\cal V}$ must vanish.

\begin{figure}[!htb]
\centerline{
{\includegraphics*[height=4.2in,width=4.5in]{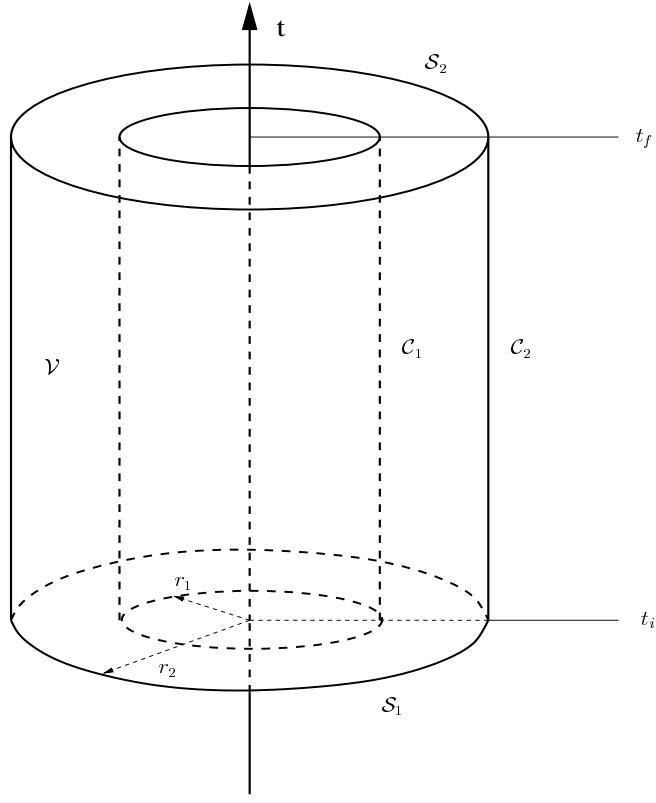}} }
\caption{Spacetime region ${\cal V}$ [see equation (\ref{volume})]
where the conservation law (\ref{elaw}) is tested.
\label{intdomain}}
\end{figure}

The region ${\cal V}$ is chosen as follows (see Figure~\ref{intdomain}):
It consists of two cylinders, ${\cal C}_1$ and ${\cal C}_2$, that extend
along the timelike Killing direction, and two slices, ${\cal S}_1$ and 
${\cal S}_2$, orthogonal to the timelike Killing, such that they cut the
cylinders forming a closed 2+1 spacetime region.  Then, the spacetime region
${\cal V}$ can be described as 
\begin{equation}
{\cal V} \equiv \left\{ (x^\mu)=(t,x,y) ~|~ t_f > t > t_i\,,
r_2 > r > r_1\,, 2\pi > \theta > 0 \right\}\,,  \label{volume}
\end{equation}
where the angle $\theta$ is defined by $\tan\theta = y/x\,;$
$t_i$ and $t_f$ can be chosen as the initial and final
integration times respectively; $r_2$ can be taken to coincide
with the outer boundary and $r_1$ can be chosen so that
$r_1>2M$, and hence the black hole region is excluded and 
$\xi$ is timelike everywhere in ${\cal V}$.
The boundary is then given by
\begin{equation}
\partial{\cal V} = {\cal C}_1 \cup {\cal C}_2 \cup {\cal S}_1
\cup {\cal S}_2 \,,\label{boundary}
\end{equation}
where
\begin{equation}
{\cal C}_1 = \left\{ (x^\mu) ~|~ t_f > t > t_i\,,
r =  r_1\,, 2\pi > \theta > 0 \right\}\,,~~
{\cal C}_2 = \left\{ (x^\mu)~|~ t_f > t > t_i\,,
r = r_2\,, 2\pi > \theta > 0 \right\}\,,
\label{boundary1}
\end{equation}
\begin{equation}
{\cal S}_1 = \left\{ (x^\mu) ~|~ t = t_i\,,
r_2 > r > r_1\,, 2\pi > \theta > 0 \right\}\,,~~
{\cal S}_2 = \left\{ (x^\mu) ~|~ t = t_f\,,
r_2 > r > r_1\,, 2\pi > \theta > 0 \right\}\,.
\label{boundary2}
\end{equation}

When we apply the conservation law (\ref{elaw}) to our domain,
we obtain the following relation between boundary integrals:
\begin{equation}
\int_{{\cal C}_1} T^{\mu\nu}\xi_\mu\, d\Sigma_\nu +
\int_{{\cal C}_2} T^{\mu\nu}\xi_\mu\, d\Sigma_\nu +
\int_{{\cal S}_1} T^{\mu\nu}\xi_\mu\, d\Sigma_\nu +
\int_{{\cal S}_2} T^{\mu\nu}\xi_\mu\, d\Sigma_\nu = 0\,.
\end{equation}
This relation can be interpreted by saying that the difference in energy
between two given instances of time, $t_i$ and $t_f\,,$ (which corresponds
to the integrals on the spacelike slices ${\cal S}_1$ and ${\cal S}_2$)
is due to the loss of energy through the cylinders ${\cal C}_1$  
(gravitational waves being absorbed by the horizon) and ${\cal C}_2$ 
(gravitational waves escaping to infinity).

In order to evaluate these surface integrals we need first to
find the normal vector everywhere on the boundary $\partial{\cal V}$,
that is, on each of the disjoint pieces of (\ref{boundary}).
The pieces ${\cal S}_1$ and ${\cal S}_2$ [see equation~(\ref{boundary2})]
have constant time $t$,
therefore the (timelike) normals there are given by
\begin{equation}
\left. {\bm n}\right|^{}_{{\cal S}_1} =
\left. - \frac{dt}{\sqrt{1+2M/r}}\right|^{}_{{\cal S}_1}\,,~~~~
\left. {\bm n}\right|^{}_{{\cal S}_2} =
\left. \frac{dt}{\sqrt{1+2M/r}}\right|^{}_{{\cal S}_2}\,,~~~~
\end{equation}
whereas the pieces ${\cal C}_1$ and ${\cal C}_2$ [see
equation~(\ref{boundary1})] are cylinders
of constant radius $r$.  Here, it is important to remark
that both of the cylinders are assumed to be
located at $r>r_h=2M$, so that they are timelike hypersurfaces.  
In practice, we are going to take $r_1$ very close to $r_h\,.$   
Taking this into account we can write the (spacelike) normals as 
follows:
\begin{equation}
\left. {\bm n}\right|^{}_{{\cal C}_1} =
-\frac{x_idx^i}{r_1\sqrt{1-2M/r_1}}
 = -\frac{dr}{r_1\sqrt{1-2M/r_1}} \,,~~~~
\left. {\bm n}\right|^{}_{{\cal C}_2} =
\frac{x_idx^i}{r_2\sqrt{1-2M/r_2}}
 = \frac{dr}{r_2\sqrt{1-2M/r_2}}  \,.
\end{equation}

Then, the contributions of the gravitational scalar field
energy-momentum to the surface integrals are:
\begin{eqnarray}
\int_{{\cal S}_1} {}^{(\Phi)}T^{\mu\nu}\xi_\mu\, d\Sigma_\nu & = &
\frac{1}{8\pi}\int\int^{}_{{\cal S}_1} dx\,dy \left\{
\Pi^2 + (\partial_x\Phi)^2 + (\partial_y\Phi)^2 -
\frac{2M/r}{1+2M/r}\left(\frac{x}{r}\partial_x\Phi
+\frac{y}{r}\partial_y\Phi\right)^2 \nonumber \right. \\
& & \left. -\frac{4M/r}{\sqrt{1+2M/r}}
\Pi\left(\frac{x}{r}\partial_x\Phi+\frac{y}{r}\partial_y\Phi\right)
\right\} \,, \label{intphis1}
\end{eqnarray}
\begin{eqnarray}
\int_{{\cal S}_2} {}^{(\Phi)}T^{\mu\nu}\xi_\mu\, d\Sigma_\nu & = &
-\frac{1}{8\pi}\int\int^{}_{{\cal S}_2} dx\,dy \left\{
\Pi^2 + (\partial_x\Phi)^2 + (\partial_y\Phi)^2 -
\frac{2M/r}{1+2M/r}\left(\frac{x}{r}\partial_x\Phi
+\frac{y}{r}\partial_y\Phi\right)^2 \nonumber \right. \\
& & \left. -\frac{4M/r}{\sqrt{1+2M/r}}
\Pi\left(\frac{x}{r}\partial_x\Phi+\frac{y}{r}\partial_y\Phi\right)
\right\} \,, \label{intphis2}
\end{eqnarray}
\begin{eqnarray}
\int_{{\cal C}_1} {}^{(\Phi)}T^{\mu\nu}\xi_\mu\, d\Sigma_\nu & = &
-\frac{1}{4\pi}\frac{r_1}{1+2M/r_1}\int^{t_f}_{t_i}dt
\int^{2\pi}_0 d\theta  \left[ \Pi - \frac{2M/r_1}{\sqrt{1+2M/r_1}}
\left(\cos\theta\partial_x\Phi+\sin\theta\partial_y\Phi \right)\right]
\left[ -\frac{2M}{r_1}\Pi \nonumber \right. \\
& & \left. +\frac{1}{\sqrt{1+2M/r_1}}
\left(\cos\theta\partial_x\Phi+\sin\theta\partial_y\Phi \right)\right]\,,
\label{intphic1}
\end{eqnarray}
\begin{eqnarray}
\int_{{\cal C}_2} {}^{(\Phi)}T^{\mu\nu}\xi_\mu\, d\Sigma_\nu & = &
\frac{1}{4\pi}\frac{r_2}{1+2M/r_2}\int^{t_f}_{t_i}dt
\int^{2\pi}_0 d\theta  \left[ \Pi - \frac{2M/r_2}{\sqrt{1+2M/r_2}}
\left(\cos\theta\partial_x\Phi+\sin\theta\partial_y\Phi \right)\right]
\left[ -\frac{2M}{r_2}\Pi \nonumber \right. \\
& & \left. +\frac{1}{\sqrt{1+2M/r_2}}
\left(\cos\theta\partial_x\Phi+\sin\theta\partial_y\Phi \right)\right]\,.
\label{intphic2}
\end{eqnarray}
The contributions of the particle's energy-momentum to the surface
integrals are:
\begin{eqnarray}
\int_{{\cal S}_1} {}^{(\rho)}T^{\mu\nu}\xi_\mu\, d\Sigma_\nu & = &
\int\int^{}_{{\cal S}_1} dx\,dy\; \rho\mbox{e}^\Phi
\left[1-\frac{2M}{r}+\frac{2M}{r^2}(x_i v^i)\right]\left(u^t
\right)^2\,, \label{intpars1}
\end{eqnarray}
\begin{eqnarray}
\int_{{\cal S}_2} {}^{(\rho)}T^{\mu\nu}\xi_\mu\, d\Sigma_\nu & = &
-\int\int^{}_{{\cal S}_2} dx\,dy\; \rho\mbox{e}^\Phi
\left[1-\frac{2M}{r}+\frac{2M}{r^2}(x_i v^i)\right]\left(u^t
\right)^2\,. \label{intpars2}
\end{eqnarray}
Since the particle density is very small outside a small neighbourhood
around the particle location, the contribution due to the particle to the
integrals on the cylinders ${\cal C}_1$ and ${\cal C}_2$ can be neglected:
\begin{eqnarray}
\int_{{\cal C}_1} {}^{(\rho)}T^{\mu\nu}\xi_\mu\, d\Sigma_\nu \approx 0
\,, ~~~~
\int_{{\cal C}_2} {}^{(\rho)}T^{\mu\nu}\xi_\mu\, d\Sigma_\nu \approx 0
\,, \label{intparc}
\end{eqnarray}
where in (\ref{intphis1}) and (\ref{intpars1}), $\Phi$ and $\Pi$ mean
$\Phi(t_i,x,y)$ and $\Pi(t_i,x,y)$ respectively.
In (\ref{intphis2}) and (\ref{intpars2}), $\Phi$ and
$\Pi$ mean $\Phi(t_f,x,y)$ and $\Pi(t_f,x,y)$ respectively.
Both in (\ref{intphic1}) and in (\ref{intphic2}) we have used
polar coordinates instead of Cartesian ones ($x=r\cos\theta$,
$y=r\sin\theta$, $dx\,dy = r\,dr\,d\theta$).  Therefore,
in (\ref{intphic1}), $\Phi$ and $\Pi$
mean $\Phi(t,r_1\cos\theta,r_1\sin\theta)$ and $\Pi(t,r_1\cos\theta,
r_1\sin\theta)$ respectively; and in (\ref{intphic2}), $\Phi$ and $\Pi$
mean $\Phi(t,r_2\cos\theta,r_2\sin\theta)$ and $\Pi(t,r_2\cos\theta,
r_2\sin\theta)$ respectively.
In (\ref{intpars1}) and (\ref{intpars2}), the objects $\rho$,
and $u^t$ must be substituted by their expressions (\ref{source}) and
(\ref{ut2}) respectively.

Then, the conservation law that we have to test numerically is:
\begin{eqnarray}
\frac{1}{8\pi}\int\int^{}_{{\cal S}_1} dx\,dy \left\{
\Pi^2 + (\partial_x\Phi)^2 + (\partial_y\Phi)^2 -
\frac{2M/r}{1+2M/r}\left(\frac{x}{r}\partial_x\Phi
+\frac{y}{r}\partial_y\Phi\right)^2
-\frac{4M/r}{\sqrt{1+2M/r}}
\Pi\left(\frac{x}{r}\partial_x\Phi+\frac{y}{r}\partial_y\Phi\right)
\right\} \nonumber \\
-\frac{1}{8\pi}\int\int^{}_{{\cal S}_2} dx\,dy \left\{
\Pi^2 + (\partial_x\Phi)^2 + (\partial_y\Phi)^2 -
\frac{2M/r}{1+2M/r}\left(\frac{x}{r}\partial_x\Phi
+\frac{y}{r}\partial_y\Phi\right)^2 -\frac{4M/r}{\sqrt{1+2M/r}}
\Pi\left(\frac{x}{r}\partial_x\Phi+\frac{y}{r}\partial_y\Phi\right)
\right\} \nonumber \\
-\frac{1}{4\pi}\frac{r_1}{1+2M/r_1}\int^{t_f}_{t_i}dt
\int^{2\pi}_0 d\theta  \left[ \Pi - \frac{2M/r_1}{\sqrt{1+2M/r_1}}
\left(\cos\theta\partial_x\Phi+\sin\theta\partial_y\Phi \right)\right]
\left[ -\frac{2M}{r_1}\Pi
+\frac{\cos\theta\partial_x\Phi+\sin\theta\partial_y\Phi}
{\sqrt{1+2M/r_1}}\right]\nonumber \\
+\frac{1}{4\pi}\frac{r_2}{1+2M/r_2}\int^{t_f}_{t_i}dt
\int^{2\pi}_0 d\theta  \left[ \Pi - \frac{2M/r_2}{\sqrt{1+2M/r_2}}
\left(\cos\theta\partial_x\Phi+\sin\theta\partial_y\Phi \right)\right]
\left[ -\frac{2M}{r_2}\Pi +\frac{\cos\theta\partial_x\Phi
+\sin\theta\partial_y\Phi}{\sqrt{1+2M/r_2}}\right]\nonumber \\
+\int\int^{}_{{\cal S}_1} dx\,dy\; \rho\mbox{e}^\Phi
\left[1-\frac{2M}{r}+\frac{2M}{r^2}(x_i v^i)\right]\left(u^t
\right)^2 \hspace{7.5cm} \nonumber \\
-\int\int^{}_{{\cal S}_2} dx\,dy\; \rho\mbox{e}^\Phi
\left[1-\frac{2M}{r}+\frac{2M}{r^2}(x_i v^i)\right]\left(u^t
\right)^2=0\,. \hspace{7cm}  \label{claw}
\end{eqnarray}
In order to decide whether the result from the numerical computation
is satisfactory, we can normalize the previous expression with
respect, for instance, the initial energy of the particle, i.e.
the last but one line in this conservation law.

\section{The Numerical Framework\label{sec3}}
The goal of this section is twofold.  First, we want to give a brief introduction
to the Finite Element method, a numerical technique that has rarely been used in
numerical relativistic
calculations.  A detailed basic exposition of the FEM can be found in classical
textbooks like~\cite{StrFix:73,Zienkiewicz:77oc,Ciarlet_1978,Hughes:1987tj,Babuska:01bs}.
Second, we want to describe the Adaptive Finite Element method, where we present a
new local mesh refinement technique based on an interpolation error estimate
introduced recently by Chen, Sun and Xu~\cite{Chen_Sun_Xu_2003}.  This is the
technique that we investigate in this paper as a possible tool for achieving the
adaptivity that the simulations of EMRBs require.

\subsection{Introduction to the Finite Element Method\label{example}}
The FEM is a numerical technique for solving problems
described by PDEs or that can be formulated as functional minimization problems.
In what follows, we briefly introduce the basic ideas and procedures of the FEM
by using a simple example involving the following wave equation:
\begin{eqnarray}
& & (-\partial^2_t + \partial^2_x + \partial^2_y)\Psi(t,x,y) = 2\,,~~~~
t\in[0,T]\,,~~~~x\in\Omega=\{(x,y)\,|\, r=\sqrt{x^2+y^2}\leq R\}\,, \label{exeq} \\
& & \left. n^i\partial_i\Psi\right|_{\partial\Omega} = 2R\,, \label{vnbc}  \\
& & \Psi(t=0,x,y) = x^2+y^2\,,~~~~(\partial_t\Psi)(t=0,x,y)= 0\,. \label{idweq}
\end{eqnarray}
The domain $\Omega$ is a disk of radius $R$. We prescribe the von Neumann boundary
condition (\ref{vnbc}) where $n^i$ are the components of the normal to the boundary
$\partial\Omega$.  The initial data is described by (\ref{idweq}).
This problem has the simple analytic solution:
\begin{equation}
\Psi(t,x,y) = t^2+x^2+y^2\,. \label{exactsol}
\end{equation}

\begin{figure}[!htb]
\centerline{
{\includegraphics*[height=4.2in,width=4.5in]{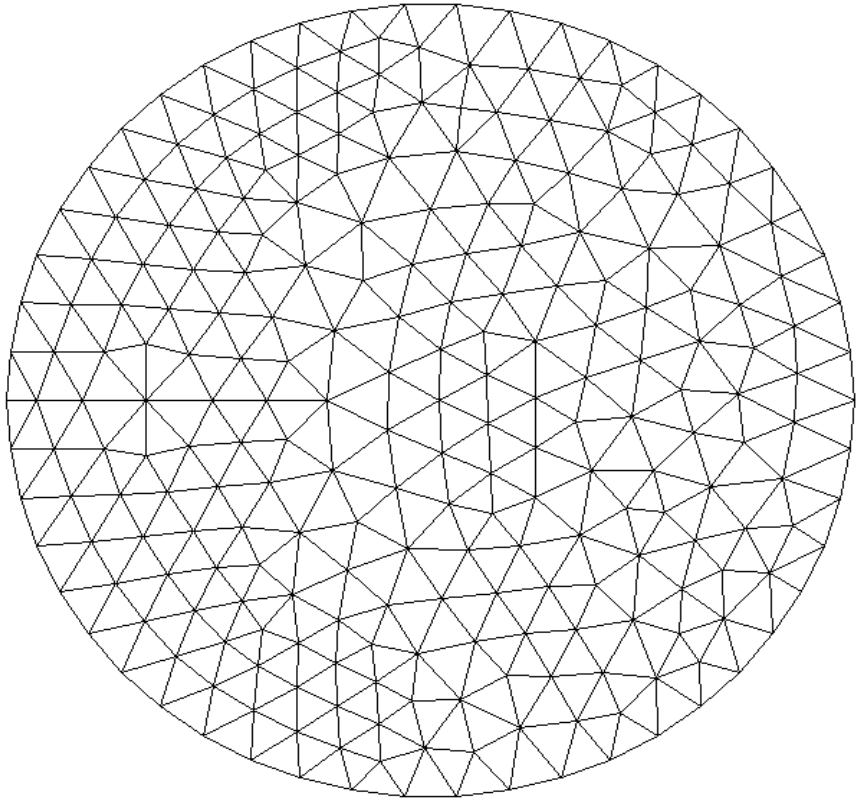}} }
\caption{Mesh corresponding to the domain $\Omega$ in Equation~(\ref{exeq}).
\label{meshexample}}
\end{figure}

We start from the discretization of the computational domain $\Omega$ into
an assembly of disjoint element domains $\{\Omega_\alpha\}$, that is
\begin{equation}
\Omega= \bigcup_\alpha\Omega_\alpha\,,~~~~~
\Omega_\beta\cap\Omega_\gamma =\emptyset~~
\mbox{for $\beta\neq \gamma$}\,.
\end{equation}
In two dimensions the element domains are typically triangles and quadrilaterals.
In this work we use only triangles.  In practice, mesh generation is carried out by
using the software included in the general-purpose software package FEPG~\cite{fpeg:2005we}.
A typical {\em mesh} (domain discretization) for our example is given in Figure~\ref{meshexample}.
Every element is equipped with a finite-dimensional functional space ${\cal F}_\alpha$,
so that we approximate our physical solution locally, at every element, as
a linear combination of functions of ${\cal F}_\alpha$ [usually, a special treatment
of the boundary elements is required].  It is very common that the functional spaces
${\cal F}_\alpha$ are formed by piecewise polynomials.   In this paper we consider
{\em linear elements}, where the element functions are first-order polynomials, i.e.
$a + b x + cy\,.$  This choice of element functions implies second-order
convergence to the solution in the $L^2$ norm.

A very important ingredient of the FEM is that it works with an integral
form of the equation we want to solve, what is called the {\em weak form} of
the problem.  To obtain it, we multiply the equation (\ref{exeq}) by a
{\em test} function $\phi$, integrate over the domain $\Omega$, and use the
Gauss or divergence theorem which introduces the boundary conditions
(\ref{vnbc}).  The result can be written in the form:
\begin{equation}
{\cal L}[\phi,\psi]\equiv
(\phi\,,\,\partial^2_t \psi) + (\nabla \phi\,,\,\nabla\psi) -
2\int_{\partial\Omega} r\,\phi\,ds + 2\, (\phi,1) = 0 \,, \label{weak}
\end{equation}
where $(,)$ denotes the inner product $(u,v)\equiv\int_\Omega u\,v\,dxdy\,,$
$\nabla$ denotes the gradient operator, and
$s$ is a coordinate on $\partial\Omega\,.$ It is important to note that the
third term is the result of introducing into the weak formulation the
boundary conditions~(\ref{vnbc}).

Using the element functional spaces we can expand our solution in terms
of {\em nodal} basis functions, $n_A(x,y)$ ($A=1,\ldots,N\,,$ being $N$ the
number of nodes), which are associated with the nodes or grid points
of the mesh.   Nodal functions use to take the unity value at the node to which
they are associated and zero at all the other nodes ($n_A(x_B)=\delta_{AB}$ for
any node $x_B$).  Since we are going to
produce a FEM discretization only in space, the expansion of our solution
in terms of the nodal functions can be written in the form:
\begin{equation}
\psi_h \in H^2(\Omega)\,,~~~~
\psi_h(t,x,y) = \sum_B \psi_B(t)\,n_B(x,y) \,, \label{femexpan}
\end{equation}
where $H^2(\Omega)$ is the Sobolev space of functions on $\Omega$ that are,
together with their first and second generalized spatial derivatives, square integrable,
that is, they belong to $L^2(\Omega)\,.$  We are assuming that $\psi_h$ belongs to
$H^2(\Omega)$ for any time $t\in [0\,, T]\,.$  The subscript $h$ denotes a scale
associated with the domain discretization, for instance it may be proportional to the
square root of the average area of the elements that compose the mesh.
In our example $h$ refers to the maximum mesh diameter in the whole domain.
An obvious property of $h$, is that the size of the elements goes to zero as
$h$ goes to zero.

In a time-dependent problem like the one we are considering, the
unknowns are going to be the functions $\psi_A(t)\,.$  The equations for these
functions are obtained from the spatial discretization.  In a Galerkin-type
formulation of the FEM, the discretized equations come from the imposition
of the vanishing of the residuals:
\begin{equation}
{\cal E}_A = {\cal L}[n_A,\psi_h] = 0 \,,  \label{residuals}
\end{equation}
which consists in taking $\phi = n_A$ in (\ref{weak}).  Introducing also
the expansion (\ref{femexpan}) yields the following system of equations
for the functions $\psi_A(t)\,.$
\begin{equation}
\sum_B M_{AB}\ddot{\psi}_B + \sum_B K_{AB}\psi_B = F_A \,, \label{femdiscre}
\end{equation}
where the matrix $M_{AB} = (n_A,n_B)$, the so-called {\em mass} matrix, and
the matrix $K_{AB} = (\nabla n_A,\nabla n_B)$, the so-called {\em stiffness}
matrix, are symmetric and positive-definite matrices.  The vector $F_A =
-2(n_A,1) + 2\int_{\partial\Omega}rn_A ds$ is sometimes called the {\em
force} vector.   Equation (\ref{femdiscre}) is the outcome of the FEM
spatial discretization, which is sometimes called the {\em semi-discrete}
form because it consist of a linear system of second-order ordinary
differential equations in time.   They are usually solved by using
Finite-Differences methods.  One of the most popular methods for second-order
in time equations is the Newmark method, which is second-order accurate in
time (see, e.g.~\cite{Hughes:1987tj}, for details).

Before we discuss the numerical implementation it is worth mentioning two
important features of the FEM:
(i) The piecewise approximations (piecewise linear in our example) of physical
fields on finite elements provide good precision even with simple approximating
functions.  Increasing the number of elements we
can achieve the desired precision.
(ii) The local character of the approximation leads to sparse systems
of equations once the problem is discretized.  This helps considerably
to solve problems with a very large number of nodal unknowns.

\begin{figure}[htbp]
\centerline{
\parbox{3.in}{\includegraphics*[width=2.5in]{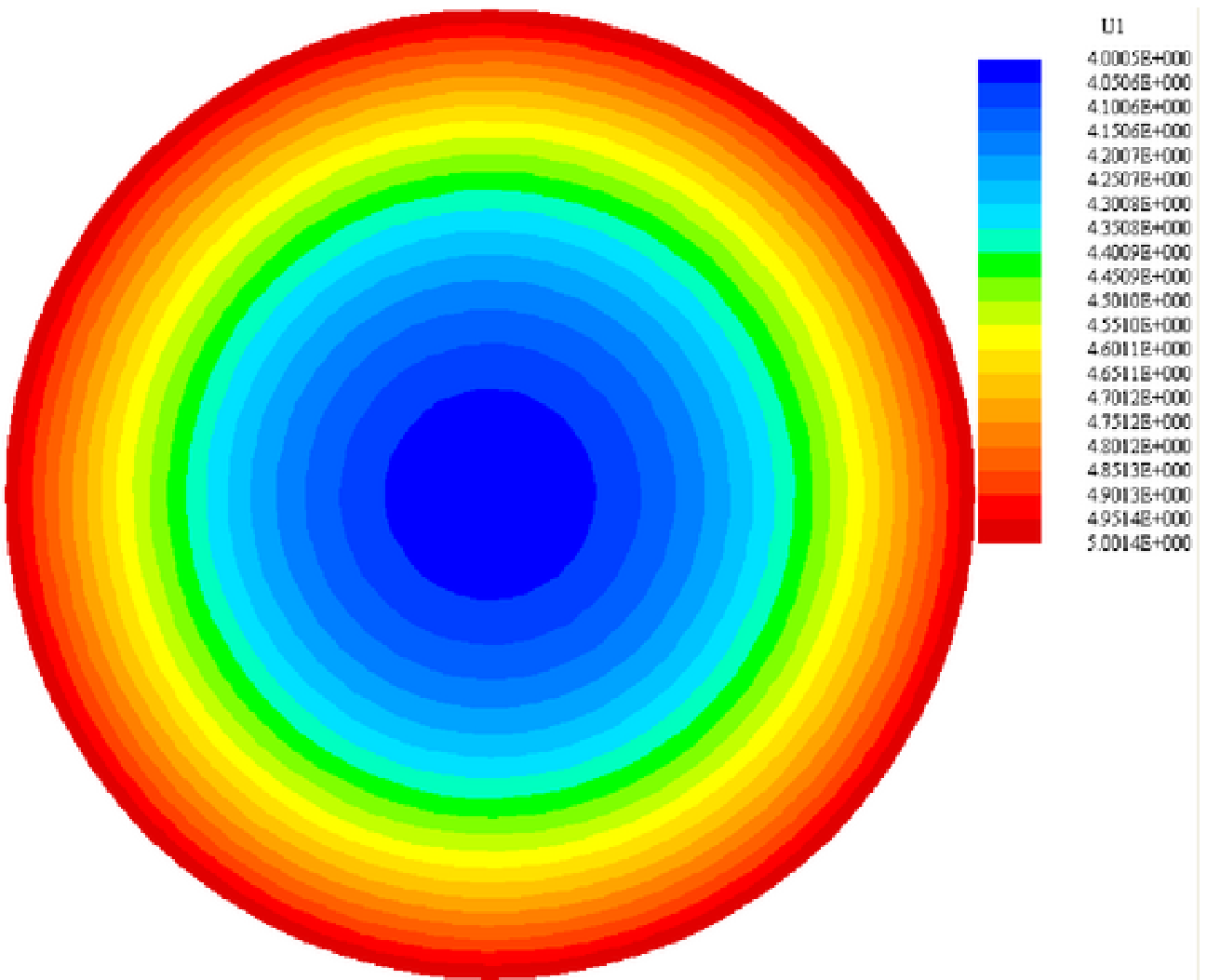}}
\parbox{3.in}{\includegraphics*[width=2.5in]{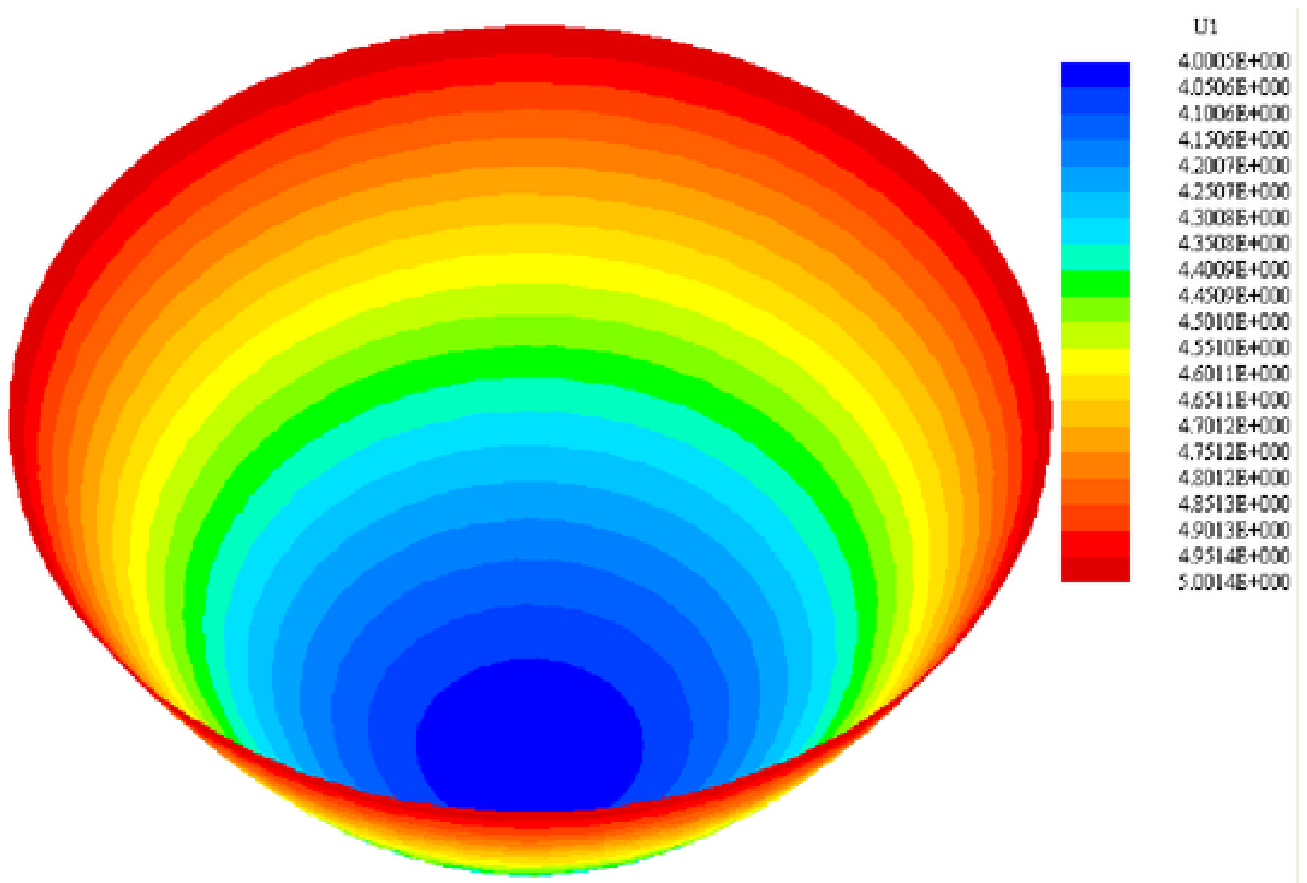}} }
\caption{Snapshots of the evolution of the solution of the example
described by equations~(\ref{exeq})-(\ref{idweq}). \label{examplesnapshot}}
\end{figure}

\begin{figure}[htbp]
\centerline{
\parbox{3.in}{\includegraphics*[width=3.1in]{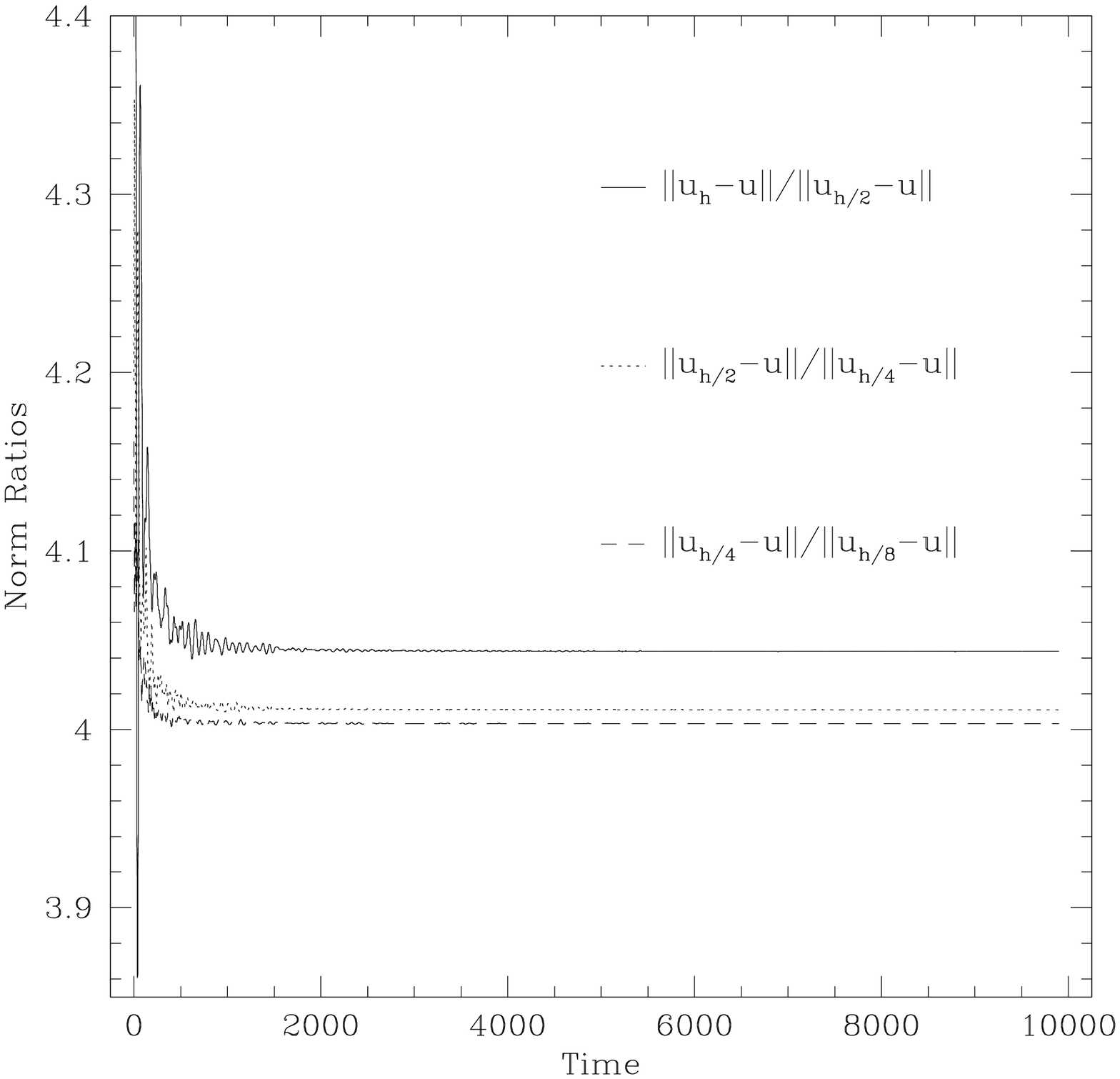}}
\parbox{3.in}{\includegraphics*[width=3.1in]{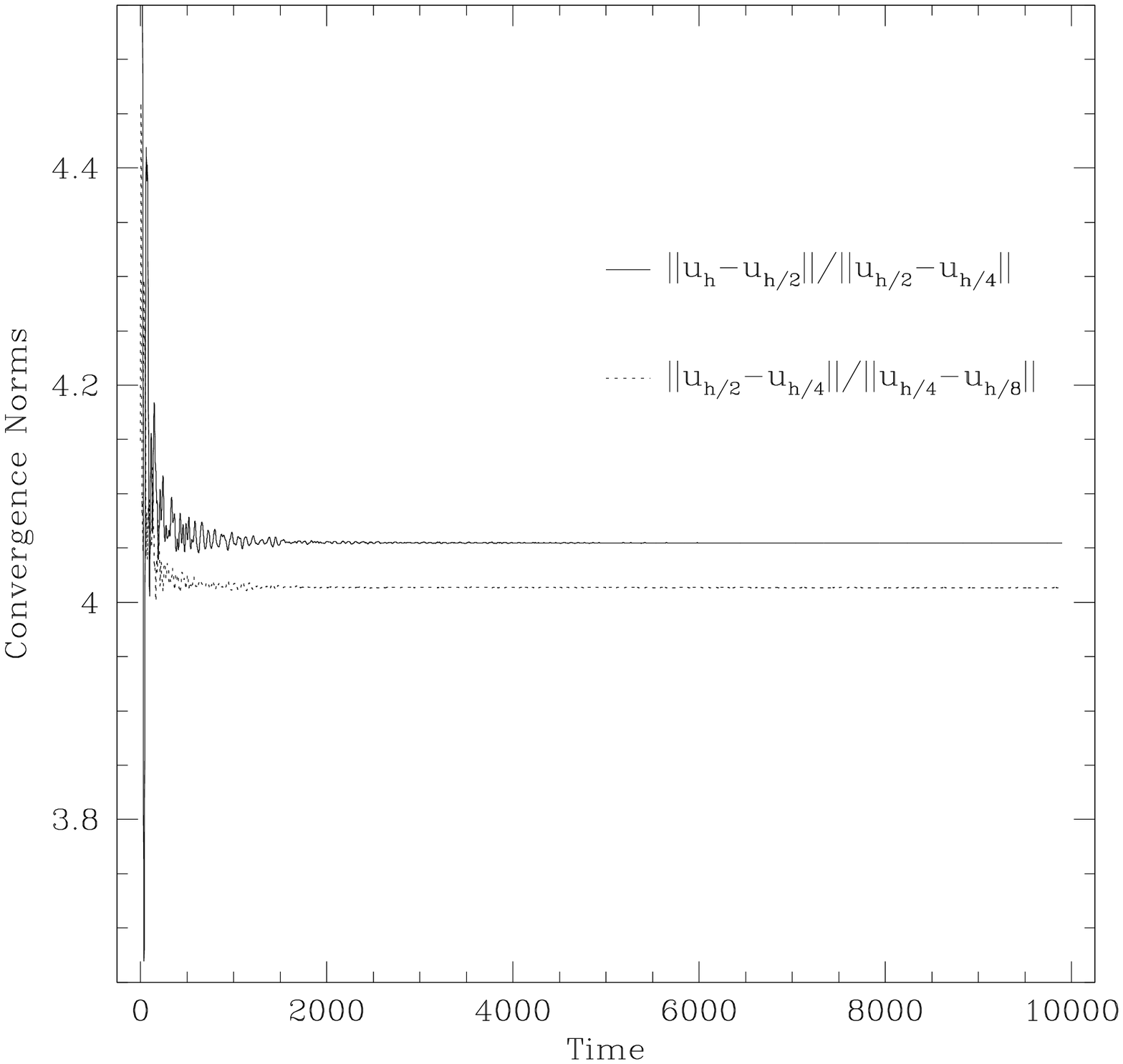}} }
\caption{Convergence properties of the solution of the
problem~(\ref{exeq})-(\ref{idweq}). On the left we show the
convergence to the exact solution~(\ref{exactsol}).  On the
right we show the $L^2$ convergence. \label{graphsconvergence}}
\end{figure}

The numerical implementation of the equations of this paper has been carried
out by using the general-purpose software package FEPG~\cite{fpeg:2005we}, which
can automatically generate finite element Fortran source code based on
component programming.  It can handle many types of problems, including
time-dependent non-linear ones, like the one we are interested in.
As a test, we have implemented the example described in this section
in FEPG (see a snapshot of the evolution in Figure~\ref{examplesnapshot}).
We have also studied the convergence properties of the solution.
Here, it is important to point out that in an unstructured mesh,
like the one we are using in this example, to perform a convergence test is
not as straightforward as it is for structured meshes. We have to define
properly the scale $h$ such that by changing it we obtain the correct
convergence.
Starting from the initial mesh (see Figure~\ref{meshexample}), we call the 
solution we obtain $\psi_h$.  Then, we globally refine this initial
mesh by transforming every initial triangular element into four smaller
triangular elements by connecting the three mid points of each edge.
Solving our example equation on this mesh leads to a more accurate
solution that we call $\psi_{h/2}$, and whose associated scale is
$h/2$.  By repeating this refinement process we get finer meshes,
and by solving our equation on them we obtain more accurate
solutions, $\psi_{h/2^k}$, with associated scale $h/2^k\,.$
We have checked that the solution we obtain converges quadratically
in the scale $h$ to the exact solution~(\ref{exactsol}) by studying
the norms $\|\psi_{h/2^k} - \psi_{\mbox{\tiny exact}}\|_{L^2}$ (see the
left of Figure~\ref{graphsconvergence}).  We have also checked that it 
convergences quadratically in the usual way, without making use of the 
exact solution, just by comparing the norms  $\|\psi_h - \psi_{h/2}\|_{L^2}$ 
and $\|\psi_{h/2} - \psi_{h/4}\|_{L^2}$ (see the right of 
Figure~\ref{graphsconvergence}).

\subsection{Finite Element Discretizations of the Toy Model}
In this section, we are going to develop a FEM formulation for
our toy model in the spirit of the ideas presented above.
The ingredients of this problem are the following:
(i) The computational domain was described in subsection~\ref{toysu}
and shown in Figure~\ref{domain}.  (ii) The equations that describe
our model are the PDEs given by equations~(\ref{fe1},\ref{fe2}).
(iii) The boundary conditions and initial data are described in 
subsection~\ref{toysu}.

\begin{figure}[!htb]
\centerline{
{\includegraphics*[height=4.2in,width=4.5in]{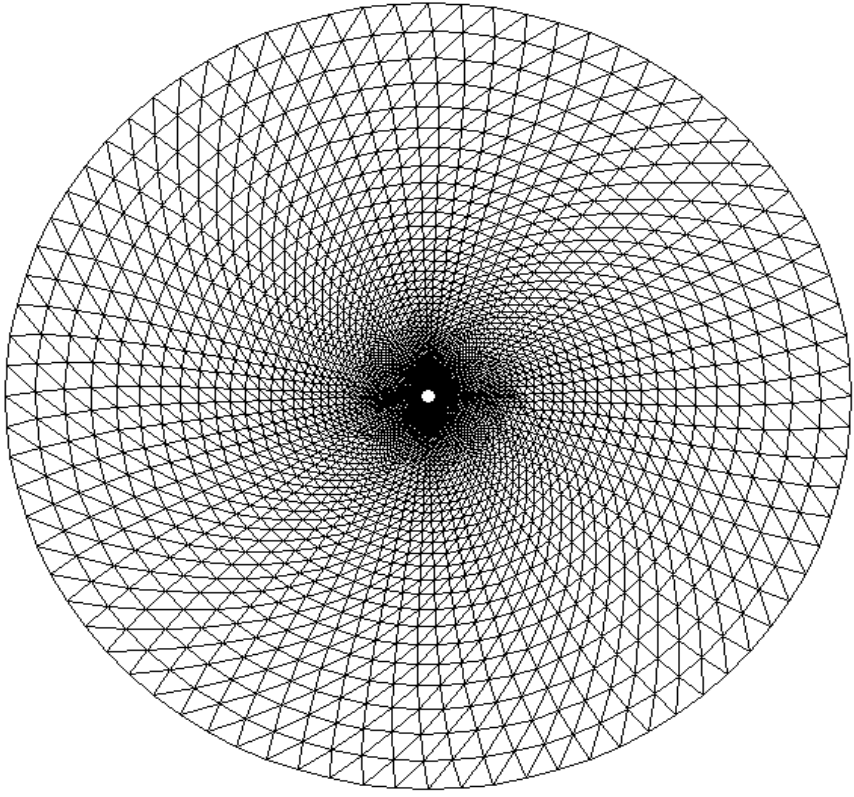}} }
\caption{Triangularization of the spatial domain given in Figure~\ref{domain}.
\label{triangularization}}
\end{figure}

We start from the discretization of the computational domain 
(see Fig.~\ref{domain}).  We use linear triangular elements.
The aspect of the resulting triangularization is shown in 
Fig.~\ref{triangularization}.  It is worth pointing out that
the FEM is specially well suited for complex domains.  In our
case, this allows us to use circular boundaries, which adapt
better to the characteristics of our problem.  This is specially
important in the case of the inner boundary, which corresponds 
to the fact that we have excised the black hole singularity from
the computational domain.

The next step in our development is the finite element discretization of the 
equations~(\ref{fe1},\ref{fe2}).  We need to construct a {\em weak} 
formulation of these equations.  To that end, it is very convenient to rewrite
the equations in the following equivalent form:
\begin{eqnarray}
\alpha^{-1}\partial_t\,\Phi &=&
\alpha^{-1}\beta^i\,\partial_i\,\Phi\,+ \Pi \,,  \label{mfe1}\\
\alpha^{-2}\partial_t\,\Pi &=&
\alpha^{-2}\beta^i\,\partial_i\,\Pi+
\partial_i(\sqrt{h}\,h^{ij}\partial_j\Phi) + \alpha^{-1}a^i \,
\partial_i\Phi + \alpha^{-1} K\Pi\ -
4\,\pi\alpha^{-1}\,e^\Phi\,\rho \,, \label{mfe2}
\end{eqnarray}
where we have used the fact that $\alpha\sqrt{h}=1\,.$
The main reason for casting the equations in this form is that it 
brings the second-derivative terms into the form of the divergence of 
a spatial vector, without any additional factors.
As we did with the example of the previous subsection, we are going to 
discretize the system of PDEs~(\ref{mfe1},\ref{mfe2}) for the unknowns 
$\Phi$ and $\Pi$ by using a FEM discretization (a Galerkin-type formulation) 
for the spatial dimensions and by using Finite Differences methods in time.  
In this sense, it is important to take into account that our
system of equations is of first order with respect to time derivatives 
and of second order with respect to spatial derivatives.  
In order to discretize these equations, particular attention has 
to be paid to the convection terms in order to keep them under control.  
We deal with this issue by including additional artificial viscosity 
into the finite element equations that we obtain.

For the FEM spatial discretization the finite element space that we use
is $S_h\subset H^2(\Omega)$,  which consists of piecewise triangular 
linear interpolation functions.  Then, taking all this into account
and operating in a similar way as we did in the example of 
subsection~\ref{example},  the discretized problem that we obtain can be 
introduced in the following way: We need to find 
$(\Phi^{n+1}\,, \Pi^{n+1})\in S_h\subset H^2(\Omega)\,,$ 
such that the equations
\begin{equation}
(\alpha^{-1}\Phi^{n+1},\phi) =
(\alpha^{-1}\Phi^{n},\phi)+(\Delta t\,\frac{\beta^i}{\alpha}
\partial_i\Phi^{n},\phi)-\delta(h)(\Delta t\,\frac{\beta^i}{\alpha}
\partial_i\Phi^{n},\Delta t\,\frac{\beta^i}{\alpha}\partial_i\phi\,)
+ (\Delta t\,\Pi^{n},\phi)\,,  \label{femfe1}
\end{equation}
\begin{eqnarray}
(\alpha^{-2}\Pi^{n+1},\varphi) &=&
(\alpha^{-2}\Pi^{n},\varphi) +
(\Delta t\,\frac{\beta^i}{\alpha^2}\partial_i\Pi^{n},\varphi) -
\delta(h)(\Delta t\,\frac{\beta^i}{\alpha^2}\partial_i\Pi^{n},
          \Delta t\,\frac{\beta^i}{\alpha^2}\partial_i\varphi) -
\Delta t\,(\sqrt{h}\,h^{ij}\partial_j\Phi^{n+1},\partial_i\varphi)
{\nonumber}\\
& + & \Delta t\,(\frac{a^i}{\alpha}\partial_i\Phi^{n+1},\varphi)
  +   \Delta t\,(\alpha^{-1}K\Pi^{n}\,\varphi)
  -   \Delta t\,(4\pi\alpha^{-1}e^{\Phi^{n+1}}\,\rho,\varphi)
{\nonumber}\\
& - & \int_{\partial\Omega_{out}}\frac{\alpha^{-1}}{(1+\frac{2M}{r})}
(\Phi^{n+1}-\Phi^{n}+\frac{\Phi^{n+1}}{2r}\Delta t)\varphi\,ds
-\Delta t\int_{\partial\Omega_{in}}\frac{\alpha^{-1}}{(1+\frac{2M}{r})}
\frac{x^i}{r}\partial_i\Phi^{n+1}\varphi\,ds\,, \label{femfe2}
\end{eqnarray}
hold for any $(\phi,\varphi)\in S_h$ and for $n=0,1,\ldots, N\,,$
$N\Delta t = T\,,$ where $T = t_f - t_o$ is the total computational time.
$\Delta t$ is the time-step size we use to evolve $\Phi$ and $\Pi\,.$
We use the notation $(\cdot,\cdot)$ to denote the inner product in $S_h$,
which is defined as $(u,v)=\int_\Omega uv d\Omega~~~(\forall~u,v\in S_h)$.
Finally, to deal with the {\em convection} terms, arising from the derivatives
of the fields along the shift vector, we use a streamline diffusion scheme
that includes artificial viscosity and which is specially adapted to the FEM.
The factor $\delta(h)$ is the penalty ratio of artificial viscosity, which
of course depends on the mesh size $h$.   It has to be tuned properly in order to
obtain a stable computation and at the same time an accurate solution.

The last terms of (\ref{femfe2}) consist of integrals on the boundaries
$\partial\Omega_{in}$ and $\partial\Omega_{out}$.   The way they come into
play is the following: In the case of the outer boundary $\partial\Omega_{out}$,
we obtain the last but one integral in (\ref{femfe2}) after integration by parts
of the term with second-order spatial derivatives in~(\ref{mfe2}) and imposition 
of the outgoing boundary condition~(\ref{bdc}).  In the case of the inner boundary 
$\partial\Omega_{inn}$, we obtain the last integral in (\ref{femfe2}) from the
same integration by parts.  However, the resulting integral does not correspond
to a proper boundary condition. As we have already mentioned before, we do not need 
any boundary condition at the inner boundary due the particular structure of the 
characteristics there (remember that the inner boundary is inside the horizon),
which all point inwards.  Therefore, the resulting integral is just an integration
of a term proportional to $\partial_i\Phi^{n+1}$ on the inner boundary.

Equations (\ref{femfe1},\ref{femfe2}) are the basis of the computational
procedure that we follow to obtain the solution of our problem. The
basic algorithm consists in computing first $\Phi_{n+1}$ from (\ref{femfe1}) in terms 
of $\Phi_{n}$ and $\Pi_{n}$.   The next step is to introduce $\Phi_{n+1}$ into the
right-hand side of (\ref{femfe2}) to obtain $\Pi_{n+1}$ in terms of $\Phi_{n}$
and $\Pi_{n}$ together.   Then, $\Pi_{n+1}$ has to be used in order  to
compute the value of $\Phi$ in the next time step, and so on.   It is not 
difficult to show that this algorithm is second-order accurate both in space and
in time.

\subsection{Computing the Motion of the Particle\label{motionp}}
In the procedure we have just described we have omitted the role of the motion of 
the particle.  It is clear that in order to evaluate the right-hand side of (\ref{femfe2}) 
we need to introduce the position of the particle.  And for that, we need to solve
the set of equations (\ref{peq1},\ref{peq2}).  These ODEs have to be integrated
simultaneously with the PDEs, which means that every time we evaluate the 
right-hand of (\ref{femfe2}) we need to evolve them a time step 
$\Delta t\,.$  And then, to use use the outcome of the integration of the ODEs (new
position and velocity of the particle) in order to evaluate the source of
equation~(\ref{fe2}) [equation~(\ref{femfe2}) in the discretized system].

The type of numerical algorithm we use to solve the ODEs (\ref{peq1},\ref{peq2})
has to take into account the particular structure of the equations~(\ref{peq2}),
which are the non-trivial ones.  These equations contain two differentiated terms,
$f^i_{\mbox{\footnotesize g}}$ and $f^i_\Phi$.  The first term would give us
the geodesic motion around a Schwarzschild Black Hole, and therefore the time
scale of the changes induced by the term $f^i_{\mbox{\footnotesize g}}$ is the
orbital period of the geodesic that the particle would follow by ignoring
the radiation reaction effects.   The second term contains the gradients
of the scalar field $\Phi$ in the neighbourhood of the particle position.
These terms are the responsible, in our toy model, of the radiation reaction
effects, and therefore, the time scale of the changes they induce will be
in general much smaller than the orbital time scale.

Taking into account that these ODEs are nonlinear both in $z^i$ and in $v^i$,
we are just going to use an explicit scheme with a time step $\Delta t_s$ much
smaller than the PDE time step $\Delta t\,.$   That is,
we split each PDE time step $\Delta t$ in many ODE time steps $\Delta t_s\,,$
so that we solve equations (\ref{peq1},\ref{peq2}) in
each time step $\Delta t_s$ in order to guarantee the accuracy of the
solution.  This scheme is able to approximate the accuracy that
an implicit scheme would have provided.
The specific discretization algorithm we use in our numerical computations
to evolve the particle's position, ${\bm z}$, and velocity, ${\bm v}$, from a
PDE time step $t_n$, with values $({\bm z}^n,{\bm v}^n)\,,$ to the next time step
$t_{n+1}$, with values $({\bm z}^{n+1},{\bm v}^{n+1})$ is given by:
\begin{equation}
\begin{array}{l}
{\bm v}_0^{n+1} =  {\bm v}^n \,, \\
{\bm z}_0^{n+1} =  {\bm z}^n \,, \\
\mbox{\tt From $m=0$ to $m=M-1\,$:} \\
~~~~{\bm v}_{m+1}^{n+1} = {\bm v}^{n}+\Delta t_{s}\,\left\{
{\bm f^{}_{\mbox{\footnotesize g}}}\left[{\bm v}_{m}^{n+1},{\bm z}_{m}^{n+1}\right]
+ {\bm f^{}_{\Phi}}\left[{\bm v}_{m}^{n+1},{\bm z}_{m}^{n+1},\Pi^n,\nabla\Phi^n\right]
\right\}\,, \\
~~~~{\bm z}_{m+1}^{n+1} = {\bm z}^{n}+\Delta t_{s}\, {\bm v}_{m+1}^{n+1}\,, \\
{\bm v}^{n+1} = {\bm v}_{M}^{n+1}\,, \\
{\bm z}^{n+1} = {\bm z}_{M}^{n+1}\,, 
\end{array}  \label{particlescheme}
\end{equation}
where $\Delta t_s$ denotes the ODE time step, which is related to the PDE time step
by the relation $\Delta t_s = \Delta t/M\,.$  We choose the integer $M$ in order to 
achieve the accuracy we want.  The way in which $\nabla\Phi$ is computed is by
using the corresponding FEM piecewise polynomial expansion that follows from the
expansion in nodal functions of $\Phi\,.$

\subsection{The Adaptive Finite Element Method\label{afem}}
The AFEM is the application of the classical FEM on a series of local
adaptive meshes in order to get more
accurate numerical results with less computational cost.
Starting with a given initial coarse mesh, the adaptive mesh on each level is
generated locally and adaptively in terms of a posteriori error estimate of
the finite element solution.
Usually local refinement takes place in places where the a posteriori errors
are much bigger than elsewhere, or in other words, where the finite element
solution changes steeply.  On the other hand,
for those places in which the a posteriori errors are sufficiently
small, the local derefinement will operate in order to eliminate
extra grids because in those places the solution changes
slowly.  Thus, during the finite element computation we can
adaptively adjust the mesh density without losing the numerical
accuracy but reducing considerably the computational cost.

It is obvious that the key part of the AFEM is the a posteriori error estimate.
To have a good such estimate means that one can precisely find out in
which places the mesh should be refined or derefined without introducing much 
numerical pollution.
After this is done, the rest of the procedure is standard.
For instance, the mesh bisection and the finite element
approximation.  There is presently a number of works in the literature
on these issues (see, e.g.~\cite{Morin_Nochetto_Siebert_2002,Rivara_Venere_1995,
Rivara_Iribarren_1996,Dolejsi_1998a}).  In what follows we
introduce our own a posteriori error estimate, which is an interpolation error 
estimate. 
Then, we elaborate on our local mesh improvement techniques such as refinement,
coarsening, and the smoothing strategy, which aim to minimize the
interpolation error.

The interpolation error estimate comes from recent work by Chen, Sun, and
Xu~\cite{Chen_Sun_Xu_2003}.   This estimate can be seen as the theoretical
foundations of our adaptive mesh techniques, that is, our algorithms are aimed at
minimizing (or at least reducing) the interpolation error by iteratively
modifying the grids.  We introduce the estimate through the following
definition: ``Let $\Omega$ be a bounded domain in $\mathbb{R}^n\,.$   Given a
function $u\in \mathcal  C^2(\bar \Omega)$, we say that a symmetric positive
definite matrix $H\in  \mathbb{R}^{n\times n}$ is a majoring Hessian of $u$ if
\begin{equation}
|\xi^t(\nabla^2u)(\mathbf x)\xi|\le c_0 \xi^tH(\mathbf
x)\xi\,,\quad
(\xi\in  \mathbb{R}^n\,,~\mathbf x\in\Omega) \,,
\end{equation}
for some positive constant $c_0$''.  Here, $\xi^t$ denotes the transpose
of the vector $\xi\,.$
We then use the majoring Hessian to define a new metric
\begin{equation}
H_p=(\det H)^{-\frac{1}{2p+n}}H\,,\quad (p\geq 1)\,.
\end{equation}

There are two conditions for a triangulation $\mathcal T_N$,
where $N$ is the number of simplexes (generic elements), to be a nearly optimal
mesh in the sense of minimizing the interpolation error in the $L^p$ norm.
The first condition consists in asking the mesh to capture the high
oscillations of the Hessian metric, that is,  $H$ should not change
very much on each element.  This condition can be expressed in a
more precise way by means of the following statement: ``There exists two 
positive constants $\alpha_0$ and $\alpha_1$ such that
\begin{equation}
\alpha_0\xi^tH_\tau\xi \,\le\, \xi^tH(\mathbf x)\xi \,\le\,
\alpha_1\xi^tH_\tau\xi\,,\quad
(\xi\in  \mathbb{R}^n\,,~\mathbf x\in\Omega)\,, \label{A1}
\end{equation}
where $H_\tau$ denotes the average of $H$ over $\tau \in \mathcal  T_N$''.
The second condition demands the triangularization $\mathcal T_N$ to
be {\em quasi-uniform} under the new metric induced by $H_{p}$.
This condition can also be express in a more precise way through
the following statement: ``There exists two positive constants, $\beta_0$
and $\beta_1$, such that
\begin{equation}
\frac {\sum_{J}^{}\tilde{d}_{\tau,J}^2}{\left|\tilde{\tau}
\right|^{\frac{2}{n}}}
\,\leq\, \beta_0\quad (\forall \tau \in \mathcal  T_N) \quad
{\rm and}\quad \frac{\max_{\tau\in\mathcal T}\left|\tilde{\tau}\right|}
{\min_{\tau \in \mathcal  T}\left|\tilde{\tau}\right|} \,\leq\, \beta_1\,,
\label{A2}
\end{equation}
where $|\tilde {\tau}|$ denotes the volume of $\tau\,,$ and
$\tilde{d}_{\tau,J}$ the length of the $J$-th edge of $\tau$ under the
new metric $H_p\,.$
The first inequality in (\ref{A2}) means that each $\tau$ is
shape-regular under the metric $H_p$. The second inequality means that all
elements $\tau$ are of comparable size (also under the new metric), which
is a global condition.  This means that the mesh is more dense at the regions
where $\det H_p(\mathbf x)$ is larger.   In~\cite{Chen_Sun_Xu_2003}, it has
been proven the important result that a triangulation that satisfies
both the local condition (\ref{A1}) and the global one (\ref{A2}) yields a
good approximation.  This result can be expressed in a precise way in the
form of a theorem:
{\em ``Let $u$ be a function belonging to $\mathcal C^2(\bar {\Omega})$,
$\mathcal T_N$ a triangularization satisfying the conditions (\ref{A1}) and
(\ref{A2}), and $u_{\cal I}$ the linear finite element interpolation of $u$ based
on the triangulation $\mathcal T_N\,.$  Then, the following error estimate
holds:
\begin{equation}
\left\|u-u_{\cal I}\right\|^{}_{L^p(\Omega)} \;\le\; CN^{-\frac{2}{n}}
\left\|\sqrt[n]{\det(H)}\right\|^{}_{L^\frac{pn}{2p+n}(\Omega)} \,,
\end{equation}
for some constant $C=C(n,p,c_0,\alpha_0,\alpha_1,\beta _0,\beta
_1)$. This error estimate is optimal in the sense that for a
strictly convex (or concave) function, the above inequality holds
in a reversed direction.''}

The result expressed by this theorem is the basis of the grid adaptation
algorithms that we use in this work.  Roughly speaking, for a given
function $u$, we will adapt our grids in such a way that the conditions given
in (\ref{A1}) and (\ref{A2}) are satisfied in the best possible way.
One important remark we need to make is that the validity of the theorem
stated above allows for a few exceptions of the condition (\ref{A2})
for $p<\infty$ (see~\cite{Chen_Sun_Xu_2003} for details).  This is of
particular importance since in practice it is very difficult to
guarantee that the condition (\ref{A2}) is satisfied everywhere.

The next point is to see how the Hessian matrix of the solution can be obtained
when the linear finite element approximation is used for the discretization of
the PDEs of our problem.  Since taking piecewise second derivatives of piecewise
linear functions will give no approximation to the Hessian matrix, special
post-processing techniques need to be used in order to obtain a reasonable
Hessian matrix approximation from linear finite elements.
One of the most popular techniques is the patch recovery technique proposed
by Zienkiewicz and Zhu (ZZ)~\cite{Zienkiewicz_Zhu_1992,Zienkiewicz_Zhu_1992a},
which is based on the least squares fitting on local patches.  Results from the
application of this technique demonstrate that it is robust and efficient.
The theoretical reason why the ZZ method works is largely understood to
be related to the superconvergence phenomenon for second-order elliptic
boundary-value problems discretized on a finite element grid that has certain
local symmetry (see the works of Walhbin~\cite{Wahlbin_1995} and Babu\v{s}ka and
Strouboulis~\cite{Babuska:01bs}).  These classic superconvergence
results can be used in order to justify the effectiveness of the ZZ method.
A significant improvement of this type of analysis was introduced recently by
Bank and Xu~\cite{Bank_Xu_2002a,Bank_Xu_2002b}.
In~\cite{Bank_Xu_2002a} they give superconvergence estimates for
piecewise linear finite element approximations on quasi-uniform
triangular meshes where most pairs of triangles that share a common
edge form approximate parallelograms.  In~\cite{Bank_Xu_2002a} they also analyze
a post-processing gradient recovery scheme, showing that $Q_h\nabla u_h$,
where $Q_h$ is the global $L^2$ projection, is a superconvergent approximation
to $\nabla u$.   This result leads to a theoretical justification of the ZZ
method for such type of grids (see Xu and Zhang~\cite{JXu_ZZhang_2003} for
details).

The gradient recovery algorithm that we use in the numerical examples of
this paper is based on a new approach due to Bank and Xu~\cite{Bank_Xu_2002b},
where they use the smoothing iteration of the multigrid method to develop a
post-processing gradient recovery scheme.  This scheme proves to be very
efficient for recovering Hessian matrices.  All the methods mentioned above
can be extended to anisotropic grids with some appropriate modifications, but a
theoretical justification of such extensions is still lacking. Nevertheless,
numerical experiments have given satisfactory results.

Let us now discuss techniques to improve the mesh quality.  We define
the mesh quality of a given triangulation $\mathcal T$ in terms of the
interpolation error:
\begin{equation}
Q(\mathcal T,u,p)= \left\|u-u_{{\cal I},\mathcal T}\right\|_{L^p(\Omega)}\,,
\quad (1\,\leq\, p\,\leq\,\infty)\,.
\end{equation}
There are three main ways of improving a mesh: (i) Refinement
or coarsening through splitting or merging of
edges~\cite{RBank_ASherman_AWeiser_1983,MRivara_1984,RKornhuber_RRoitzsch_1990};
(ii) edge swapping by replacing sets of elements by other such sets
while preserving the position of the points (nodes)~\cite{Lawson_1977};
and (iii) mesh smoothing, which moves the vertexes of the mesh while
keeping the connectivity~\cite{DField_1988,RBank_RSmith_1997,
MShephard_MGeorges_1991,LFreitag_MJones_PPlassmann_1995}.
We derive those techniques by minimizing the interpolation
error in the $L^p$ norm.

\begin{figure}[!htb]
\centerline{
{\includegraphics*[height=1.4in,width=3.4in]{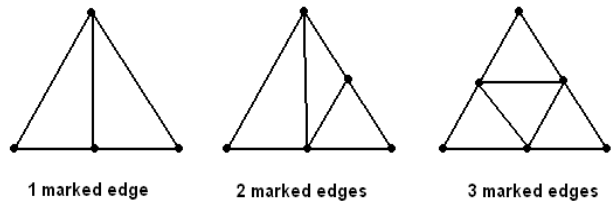}} }
\caption{Edge based refinement. \label{Fig:RefineCoarsening}}
\end{figure}

For the first method, this can be done by equidistributing
the edge lengths with respect to the new metric.
Thus, we compute the edge lengths with respect to the new metric
$H_p$ and mark edges whose length is greater than $r_1\,d$, where
$r_1\geq 1$ is a parameter and $d$ is a fixed edge length, the global average
edge length.
Then, we connect marked edges element-wise according to the
different situations that can be given.  This is illustrated
in Figure~\ref{Fig:RefineCoarsening}.
The coarsening operates like an inverse procedure to the refinement
process.  It marks the edges whose length is smaller than $r_2\,d$, where
$r_2\leq 1$ is another parameter.  We then shrink this edge to a point
and connect to the vertexes of the patch of the edge.

We consider now the case of edge swapping involving four points
$\{\mathbf x_\alpha\}$ ($\alpha=1-4$) constituting two adjacent triangles and a
convex quadrilateral.  Let $\mathcal  T_1=\triangle _{123}\cup \triangle
_{134}$ and $\mathcal T_2=\triangle _{124}\cup \triangle _{234}$ be two
triangularizations, where $\triangle _{\alpha\beta\gamma}$ stands for the
triangle made up of the points $\mathbf x_\alpha\,,$ $\mathbf x_\beta\,,$ and
$\mathbf x_\gamma\,.$   Then, we choose the triangulation
$\mathcal  T_1$ if and only if $Q(\mathcal
T_1,u,p)\leq Q(\mathcal T_2,u,p)\,,$ for some $1\leq p\leq \infty\,.$
In~\cite{LChen_JXu_2004}, we show this criteria is equivalent to
the empty circle criteria when $u(\mathbf  x)=\|\mathbf  x\|^2\,.$
Thus it is an appropriate generation of the edge swapping used in
the isotropic case to the anisotropic case.

Finally, local smoothing of the mesh adjusts the location of a vertex in its
patch $\Omega_\alpha$, which consists of all simplexes containing the vertex
$\mathbf x_\alpha$, without changing the connectivity.  The moving of a vertex
to its new location is expected to improve the quality of the mesh.
Several sweeps throughout the whole mesh can be
performed to improve the overall mesh quality. By minimizing the
interpolation error in $\Omega_\alpha$, we move the vertex $\mathbf x_\alpha$ to
the position $\mathbf  x^*$ in such a way that
\begin{equation}
\nabla u(\mathbf x^\ast)=-\frac{1}{\left|\Omega_\alpha\right|}
\sum^{}_{\tau_\beta \in \Omega_\alpha}
\Big( \nabla \left|\tau_\beta\right|
\sum_{\substack{\mathbf{x_\gamma} \in \tau_\beta \\ \mathbf{x_\gamma} \neq 
\mathbf{x_\alpha}}}
u(\mathbf{x_\gamma}) \Big) \,. \label{eq:center}
\end{equation}
The derivation of this formula can be found in~\cite{LChen_JXu_2004}.

For the application to our numerical computations we use $Q_h \nabla u_h$
and $u_h$ in (\ref{eq:center}).  Moreover, we need to make choices for the
different parameters of this adaptive mesh technique.   For the order of the
$p$ of the $L^p$ norm we use $p=1$, the $L^1$ norm. According to~\cite{LChen_JXu_2004},
the $L^1$ norm can catch singularities more efficiently than other norms.
For the multiple $r$ ($r_1,r_2$ in the text) of the global average edge length
$d$ under the new metric $H_p$ (we use $rd$ as the threshold to determine which
edge lengths under the new metric are bigger than it, we will then bisect those
edges in the local refinement process) we take $r=3$ initially.
Our local refinement procedure is a nested iteration process.  There is an outer
iteration for which, on each step, we reduce the multiple $r$ till $r_{min}$
(we take $r_{min}=1$).
The main purpose of this iteration is to resolve the singularity as precisely as
possible and to control the refinement pollution at the lowest level.
On each outer iteration step, we perform an inner iteration, and there another
number $r$ to control this iteration.  Usually we do not need this number to
establish a criterion to stop the inner iteration of local refinement.
For the sake of getting the optimal mesh, we let this iteration go until all
the edge lengths are smaller than $rd$ as before.  In practice there are occasions
when this may take too much time, and in those cases we set up a criterion as the
ones before in order to control the number of inner iterations.

\section{Numerical Results\label{sec4}}
In this section we report the results of the simulations of the toy model.
We have distinguished two types of simulations.  Those that were performed by
just using the classical FEM, without adding any extra adaptivity.  The
results are discussed in subsection~\ref{without}.  And those that were
performed by using the AFEM described in the previous section.  The results
of these simulations are discussed in subsection~\ref{with}.

\subsection{Numerical simulations of the toy model using the FEM\label{without}}
The mass ratio we have considered is $\mu = 0.01$, that is the particle's
mass is $m=0.01\,M\,.$  The inner boundary is located at $r_{in}=1\,M$,
inside the horizon $r_h=2\,M\,.$  The outer boundary has been located
at $r_{out}=50\,M$.  Regarding the initial data, the only parameters
that need to be given, in addition to the ones already given, in order
to completely specify it [see subsection~\ref{toysu}] are the width of the
Gaussian that we use to regularize the Dirac delta distribution, which we take
to be $\sigma=1\,M$, and the initial position of the particle,
which we take to be $(x_o,y_o)=(10\,M,0)\,.$ 
Finally, we give the resolution that we have used for the
simulations.  To give a measure of the spatial resolution we describe
the structure of the mesh.  It is composed of $N=9266$ triangles,
quasi-uniformly distributed, more concentrated near the center and
coarsening gradually as we approach the outer boundary.  Regarding
the resolution in the time direction, the step size we use for the time
evolution is $\Delta t\,=0.01\,M$.    The algorithm we use to study the
evolution of the gravitational scalar field $\Phi$ and the motion of
the particle is completely described by the explicit schemes of
equations (\ref{femfe1},\ref{femfe2}) and (\ref{particlescheme}).

\begin{figure}[htbp]
\centerline{ {\includegraphics*[width=3.in]{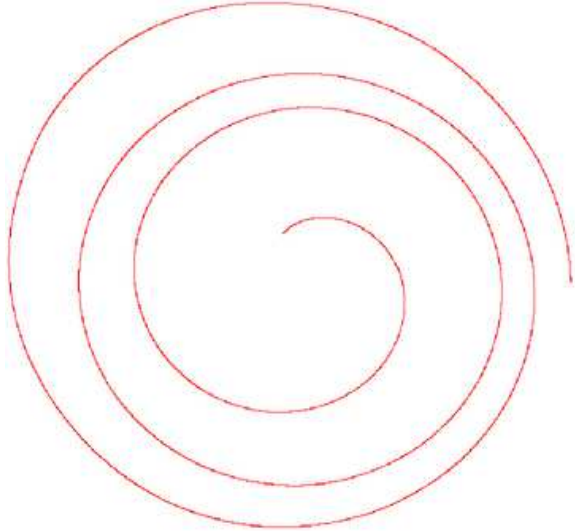}}}
\caption{Trajectory followed by the particle.\label{trajectoryfem}}
\end{figure}

\begin{figure}[htbp]
\centerline{
\parbox{3.in}{\includegraphics*[width=2.5in]{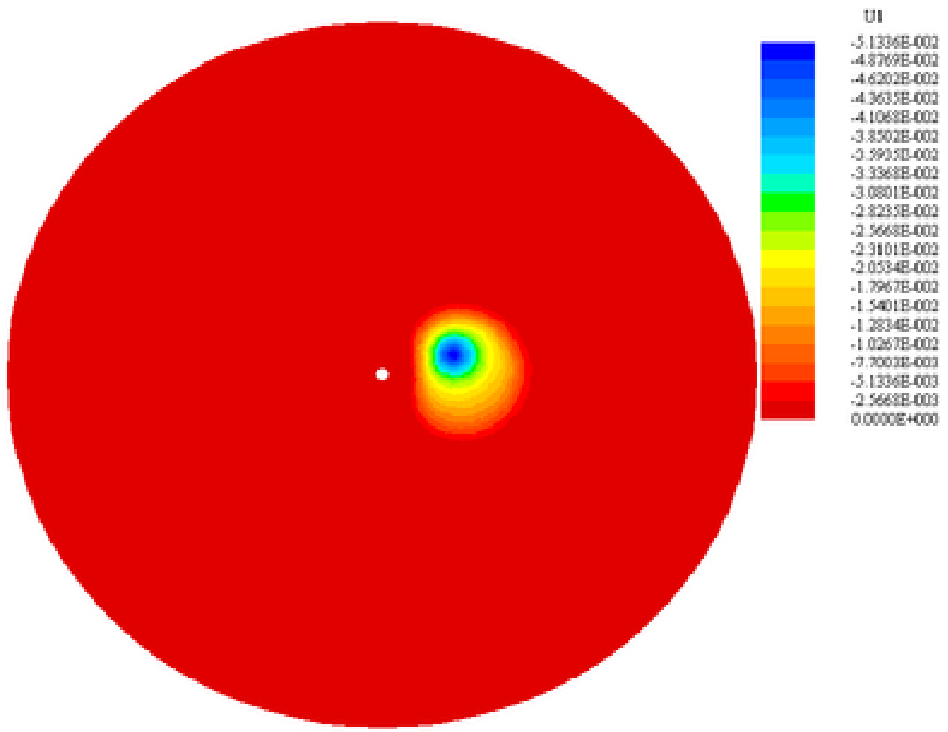}}
\parbox{3.in}{\includegraphics*[width=2.5in]{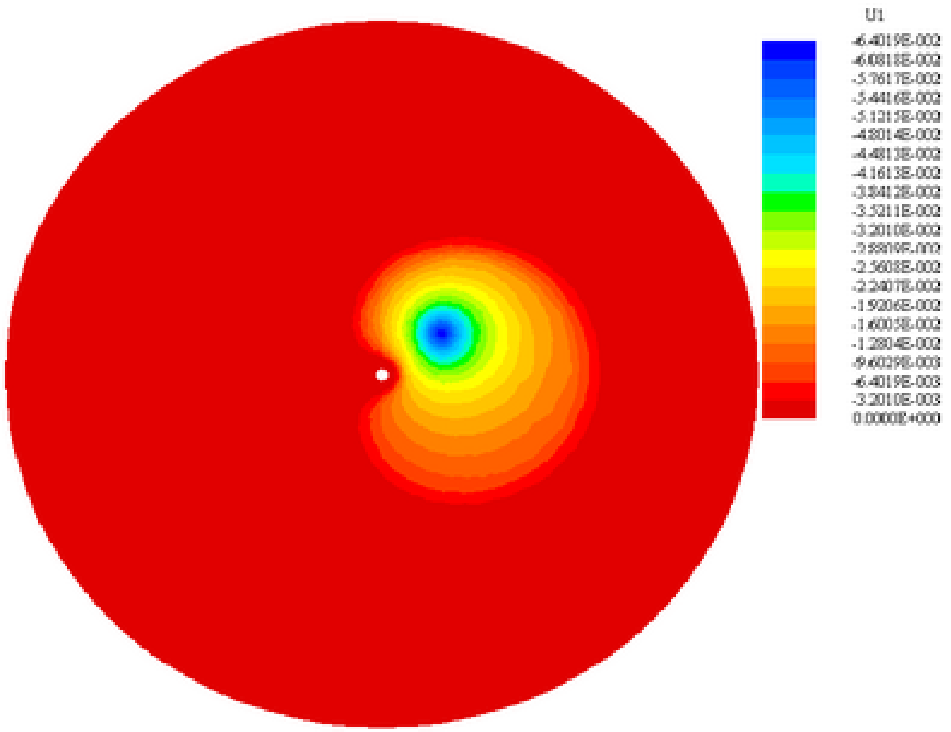}} }
\caption{Scalar gravitational field $\Phi$ at time $t=10\,M$ (n=1000) [left]
and at time $t=20\,M$ (n=2000) [right]. \label{snapshot1}}
\end{figure}

\begin{figure}[htbp]
\centerline{
\parbox{3.in}{\includegraphics*[width=2.5in]{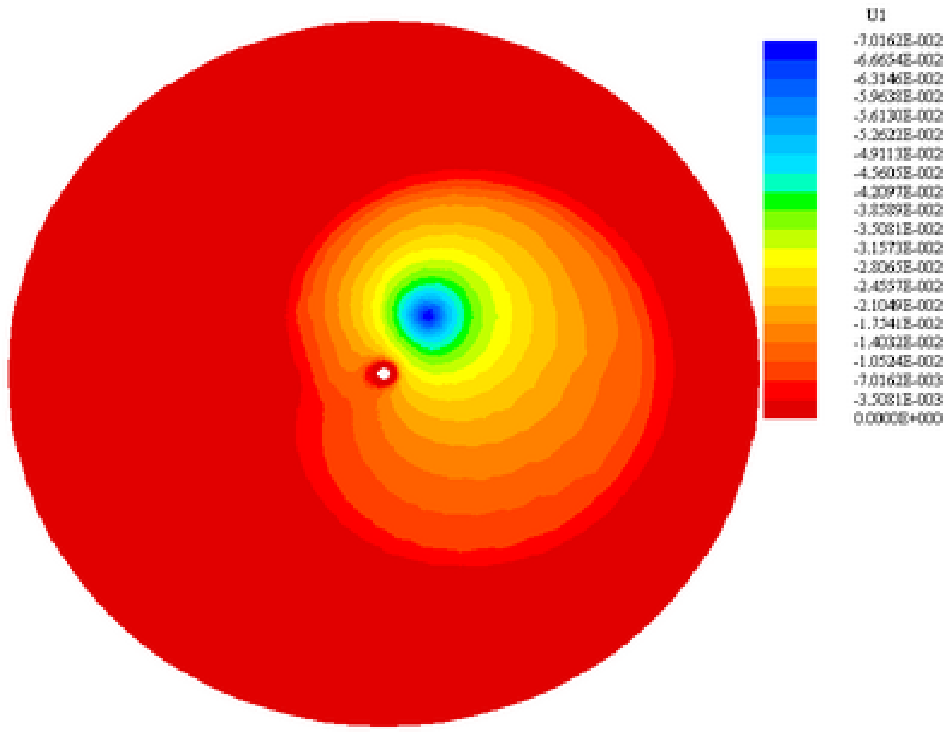}}
\parbox{3.in}{\includegraphics*[width=2.5in]{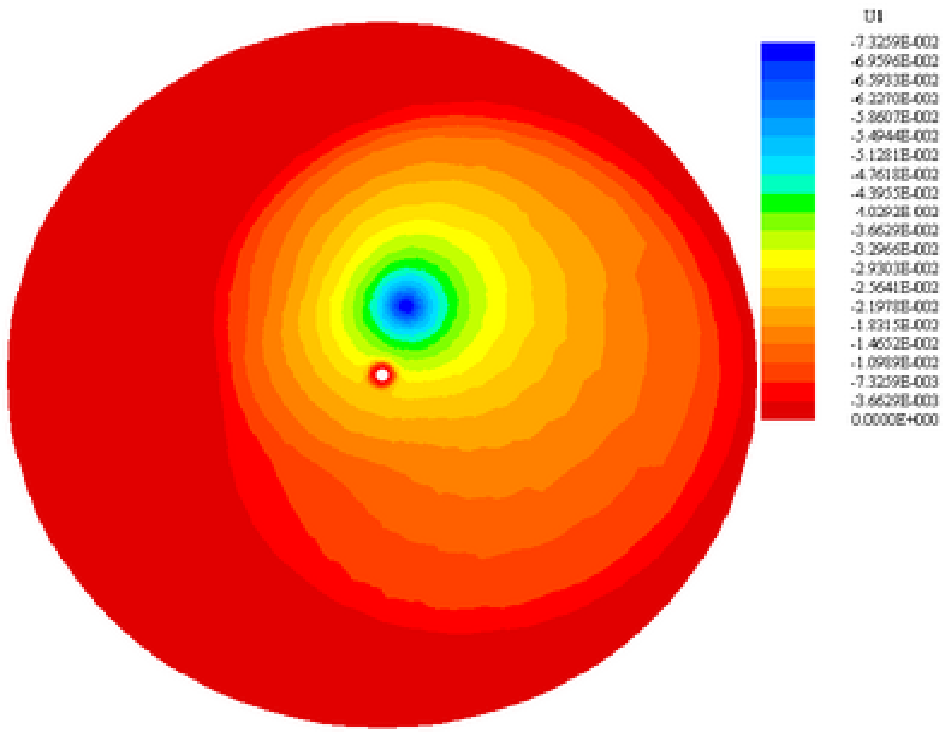}} }
\caption{Scalar gravitational field $\Phi$ at time $t=30\,M$ (n=3000) [left]
and at time $t=40\,M$ (n=4000) [right]. \label{snapshot2}}
\end{figure}

\begin{figure}[htbp]
\centerline{
\parbox{3.in}{\includegraphics*[width=2.5in]{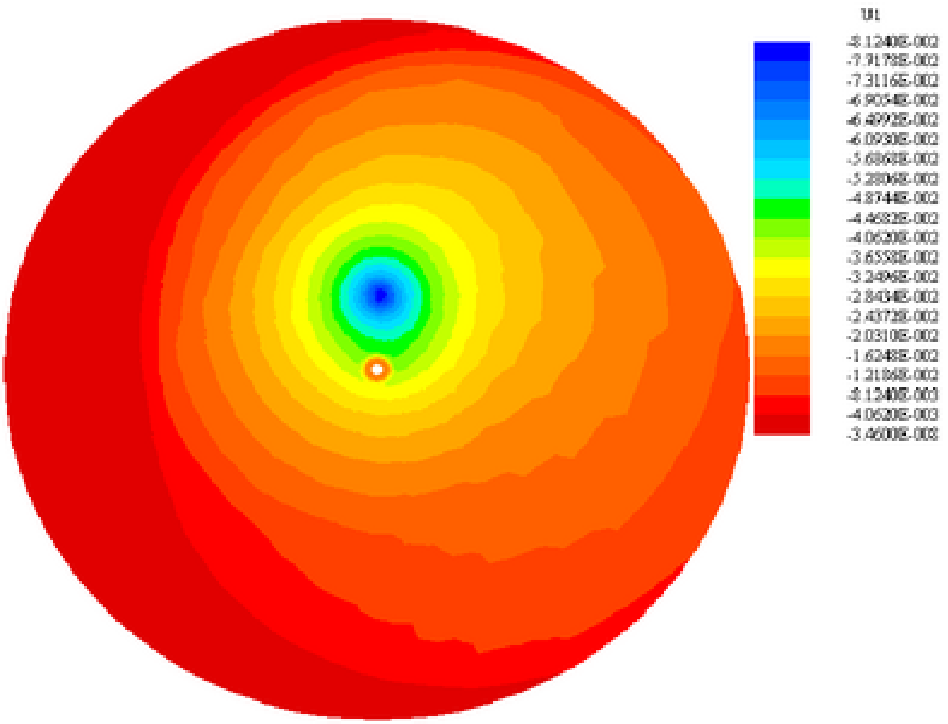}}
\parbox{3.in}{\includegraphics*[width=2.5in]{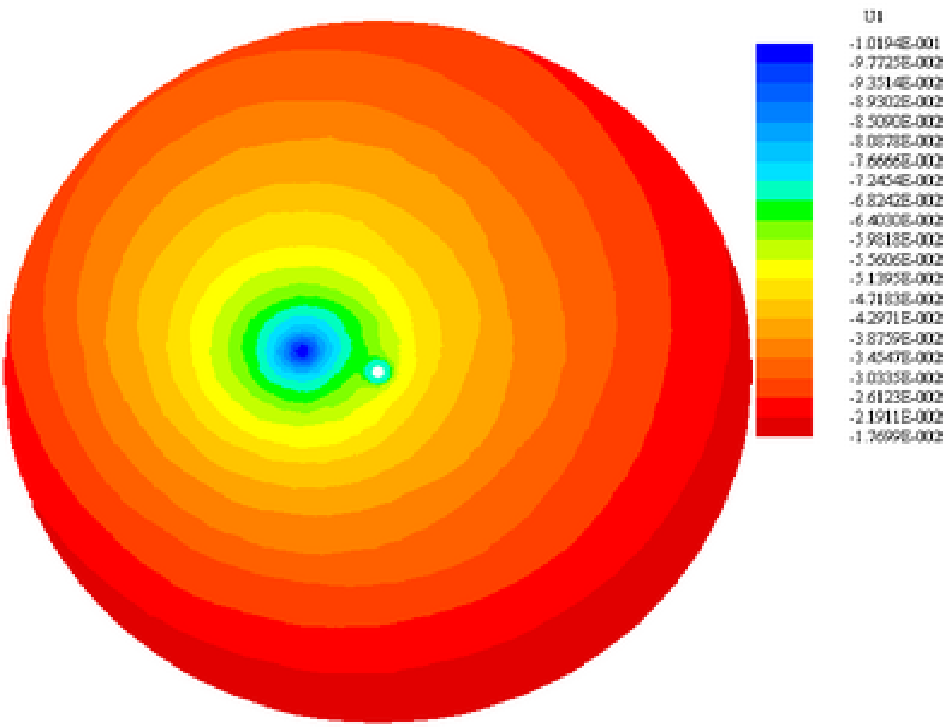}} }
\caption{Scalar gravitational field $\Phi$ at time $t=50\,M$ (n=5000) [left]
and at time $t=100\,M$ (n=10000) [right]. \label{snapshot3}}
\end{figure}

\begin{figure}[htbp]
\centerline{
\parbox{3.in}{\includegraphics*[width=2.5in]{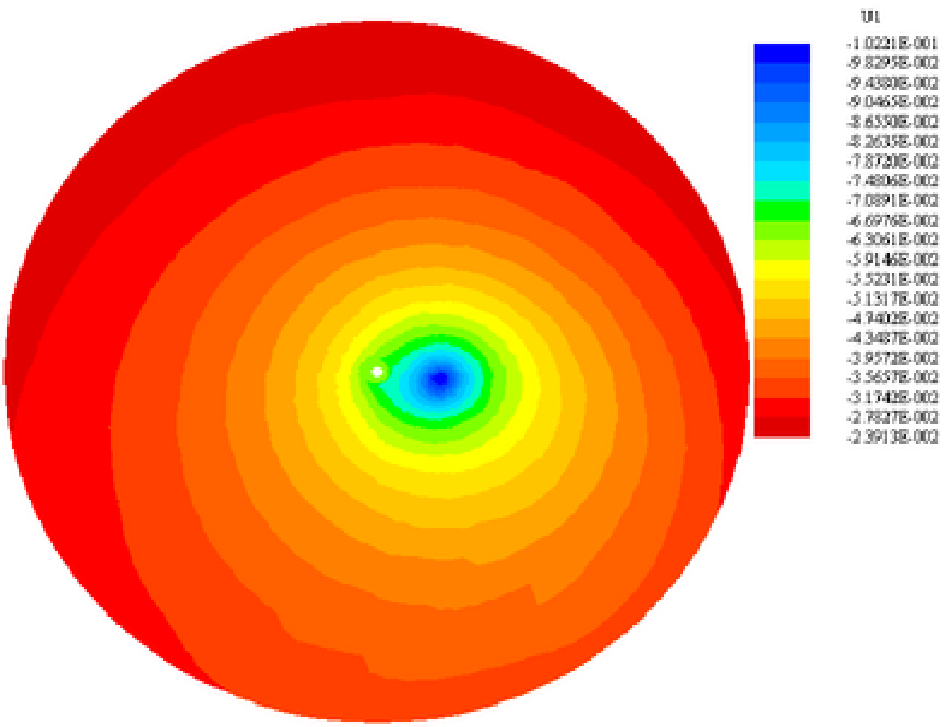}}
\parbox{3.in}{\includegraphics*[width=2.5in]{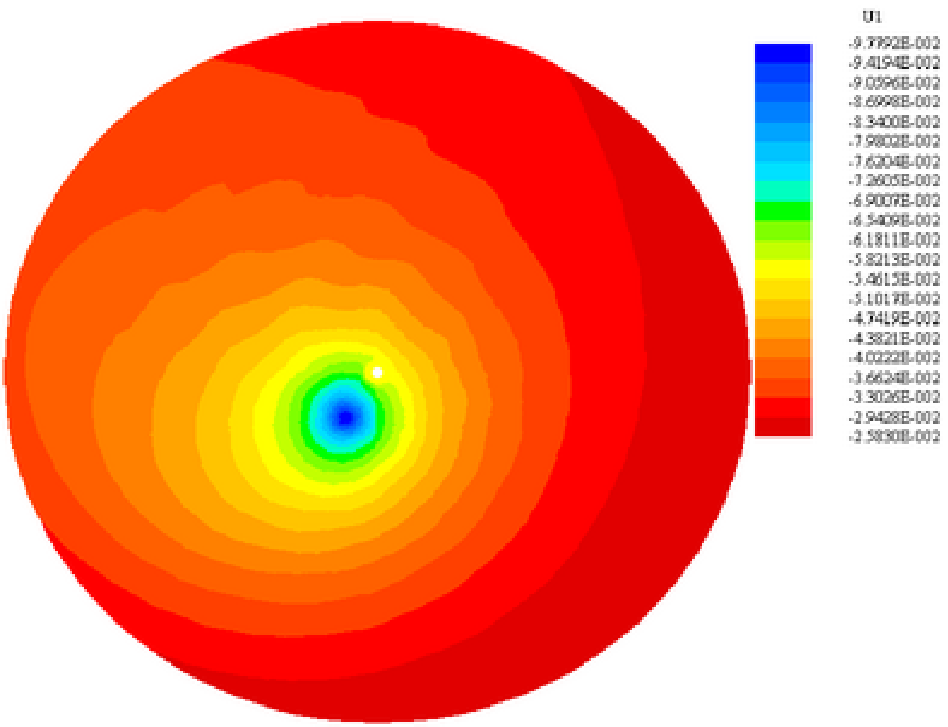}} }
\caption{Scalar gravitational field $\Phi$ at time $t=200\,M$ (n=20000) [left]
and at time $t=300\,M$ (n=30000) [right]. \label{snapshot4}}
\end{figure}

\begin{figure}[htbp]
\centerline{
\parbox{3.in}{\includegraphics*[width=2.5in]{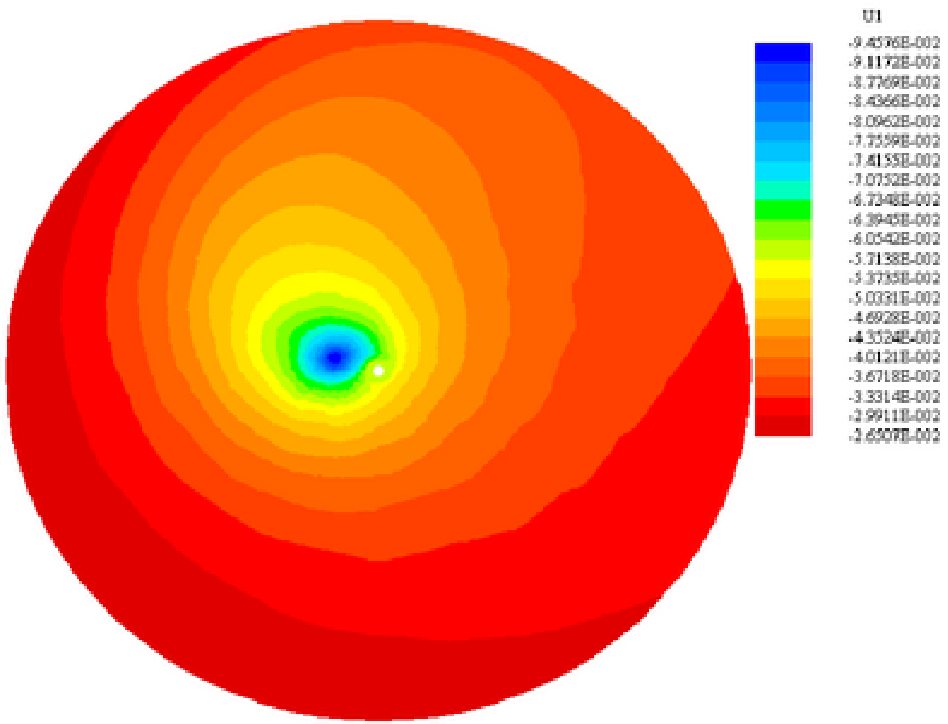}}
\parbox{3.in}{\includegraphics*[width=2.5in]{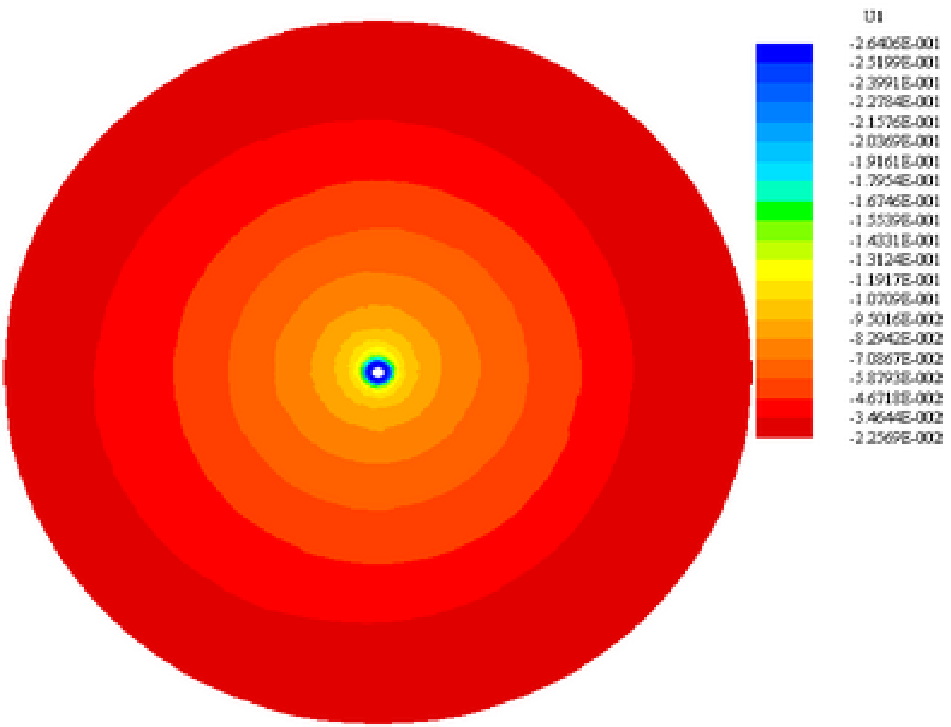}} }
\caption{Scalar gravitational field $\Phi$ at time $t=400\,M$ (n=40000) [left]
and at time $t=500\,M$ (n=50000) [right]. \label{snapshot5}}
\end{figure}

What we have observed is that the orbit of the particle (remember that
the initial data corresponds to a circular orbit in the absence of
the scalar gravitational field) shrinks gradually until the particle
reaches the horizon at $r_h=2\,M$. 
The trajectory followed by the particle is drawn in Figure~\ref{trajectoryfem}.
Snapshots characterized by the time $t$ and time-step number $n$ of the
evolution of the scalar gravitational field are shown by means of the contour
plots given in Figures~\ref{snapshot1}-\ref{snapshot5}.
The position of the particle is evident in these graphs.

\begin{figure}[!htb]
\centerline{
{\includegraphics*[width=5in]{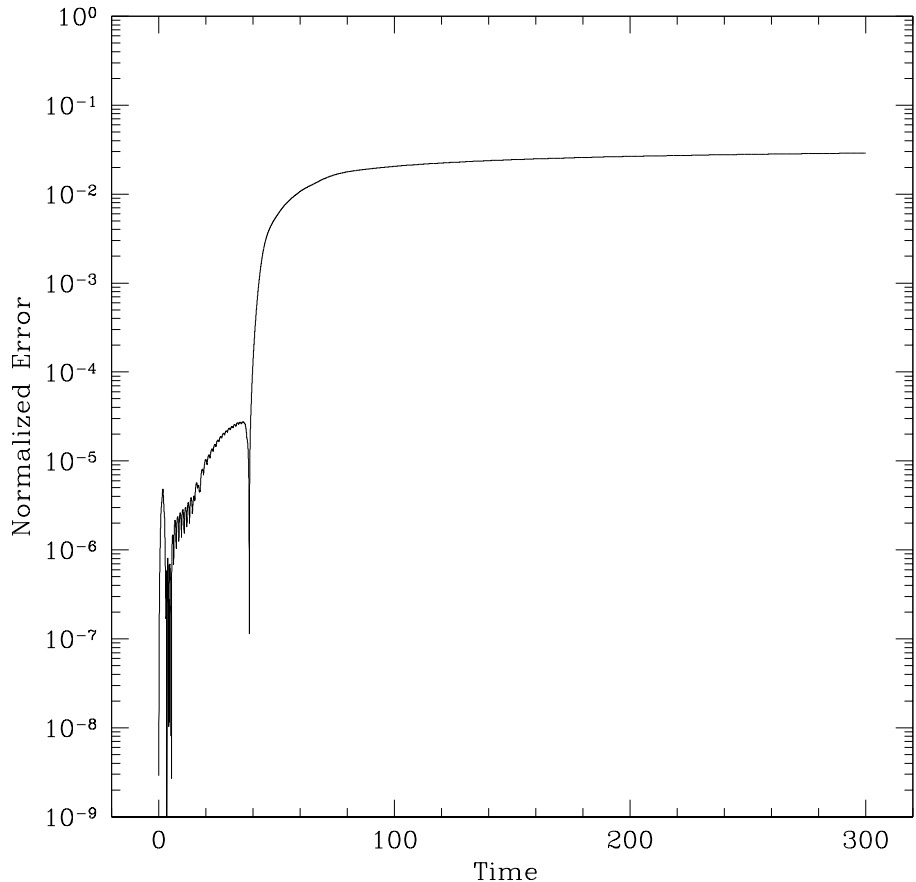}} }
\caption{Energy-Balance Test for the simulation with the FEM
(no additional adaptivity). \label{etesta}}
\end{figure}

We have checked the energy-balance law~(\ref{claw}) for the parameters
given above.  The result can be found in Figure~\ref{etesta}, where the
horizontal axis denotes the time and the vertical axis indicates the value
of the left-hand side of (\ref{claw}) in units of $m\,$.
As we can see there, the energy law~(\ref{claw}) is satisfied up to a
certain level along the evolution.  However, from $t=40\,M$ the error
in the energy-balance test grows and stabilizes around a different but bigger value.
This growth is due to outer boundary effects: The particle started
from a position $r=10\,M$ and the outer boundary is located at $r=50\,M\,.$
As we have mentioned in the discussion of the boundary conditions in
subsection~\ref{toysu}, the outer boundary condition~(\ref{bdc}) is not optimal,
and induces some error in the solution.  This could be improved by either
moving the outer boundary further out or by using an improved outer
boundary condition.

The key point in these computations is the ability of our numerical scheme
to resolve the source term describing the particle.  It enters the equations
for the scalar gravitational field $\Phi$ as a very singular source term.
This source term depends on the position of the particle, but at the
same time, the position of the particle depends of the gradients of
$\Phi\,.$  Therefore, it is very important first to compute properly
the effect of the source $\rho$ on the field $\Phi$, and then to
compute accurately the gradient of the field itself.  We have already
mentioned that in the computations shown in this subsection the
width of the Gaussian has been taken to be $\sigma=1\,M$ in order to make
the source sufficiently smooth for the resolution we have used.
In this sense,  if we try to use a smaller width, for instance
$\sigma=0.1\,M\,,$ the classical FEM will fail to provide the expected
accuracy, although it still provides a numerical solution.
Here is where one realizes the potential of the AFEM as a better
choice to carry numerical computation for a model presenting features
similar to the ones of our model.  In the next subsection we describe
how we have implemented the AFEM in the case of the toy model and
show that it provides a reasonable solution for the case in which
$\sigma=0.1\,M\,.$

\subsection{Numerical simulations of the toy model using the AFEM\label{with}}
We have just concluded that the central part of the numerical simulation
of our toy model, and more in general of EMRBs, is the proper
resolution of the matter source, $\rho\,,$ and of the gravitational field,
$\Phi\,,$ in the surroundings of the particle position, which is a key
issue in order to compute accurately the motion of the particle and
hence, the waves that are emitted as a consequence.  To that end,
within the framework we have established above, we want to able to
compute accurately for small values of the particle's width $\sigma$.
In the limit $\sigma\rightarrow 0$ we recover the distributional
description of the particle's that appears in the continuum description.
Then, it is obvious that better accuracy in the terms discussed above
means to increase the resolution around the particle's position,
and in this sense the AFEM is a natural choice since it provides
the resolution required at the different regions of the computational
domain, which maintains the computational cost at realistic levels.

According to the theoretical foundations of the adaptive mesh
technique presented above, in order to apply mesh adaptivity successfully in
our numerical computations, first of all, we have to study which
quantity in the problem we are dealing with can be used in order
to determine where and how the mesh should be refined.  In technical
terms this means that we have to be able to determine the places
where refinement (or derefinement) has to take place in terms of
the Hessian matrix of the selected quantity.  We have to look for
place where the majoring Hessian matrix is much bigger than
anywhere else, which are the places where the quantity we have
chosen changes very fast.  In the case of our toy model there is no
too much choice.   We can either take the field $\Phi$ or the source
$\rho$.   In the case of $\Phi$, it is interesting in the sense that
it is one of the goals of our computations.  However, since it is
the solution of a wave-type equation, it evolves like a wave in the
sense that the profile of the solution will have maxima and minima
which will be captured by the adaptive mesh procedure, and hence
most parts of the domain will be eventually refined, and this may
not be the most efficient way to proceed.  The other choice,
the particle's source $\rho\,,$ is a better choice for the simple reason that
we are also solving for the particle's position at the same time that
we are solving the PDEs for the field.  Therefore, the recognition of
the regions that need refinement is going to be simpler in our
problem.  That is, the most efficient way of adapting the mesh for
our computations is to concentrate on the region where the particle
is present and follow it from there.

If we look at the behaviour found in the previous subsections
(see Figures~\ref{snapshot1}-\ref{snapshot2}), it happens that the
bigger values of $\Phi$ occur around the particle's position.
Therefore, refining there is very convenient at that stage of the
computation.  A possible drawback of using the particle's source as the
refinement criterion is that the situation changes at later times
when the field $\Phi$ is present all over the computational domain
and it may be a need for refinement in other places like close to
the horizon.  However, in the numerical simulations we have performed
we find that using the source term $\rho$ for refinement is much better than
using the field $\Phi\,.$  By calculating the Hessian matrix of the
particle's source we can exactly know where and how to get local mesh
refinement on the particle.

\begin{figure}[htbp]
\centerline{ {\includegraphics*[width=3.in]{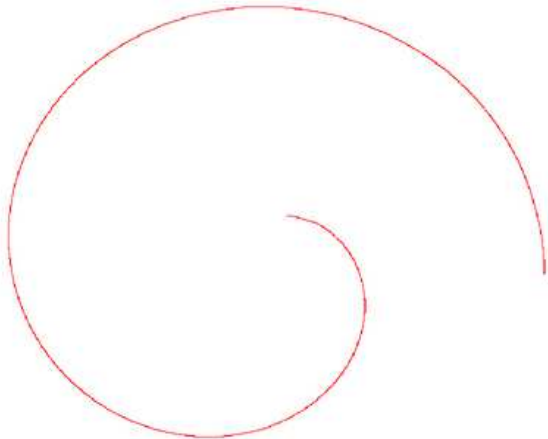}}}
\caption{Trajectory followed by the particle.\label{trajectoryafem}}
\end{figure}

\begin{figure}[!htb]
\centerline{
\parbox{3.in}{\includegraphics*[width=2.5in]{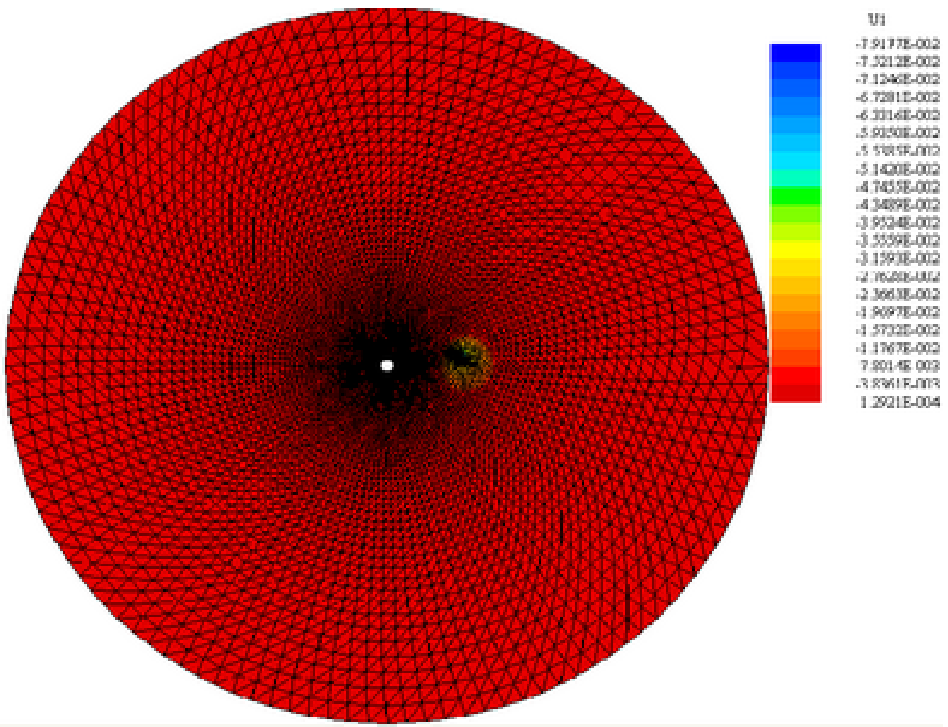}}
\parbox{3.in}{\includegraphics*[width=2.5in]{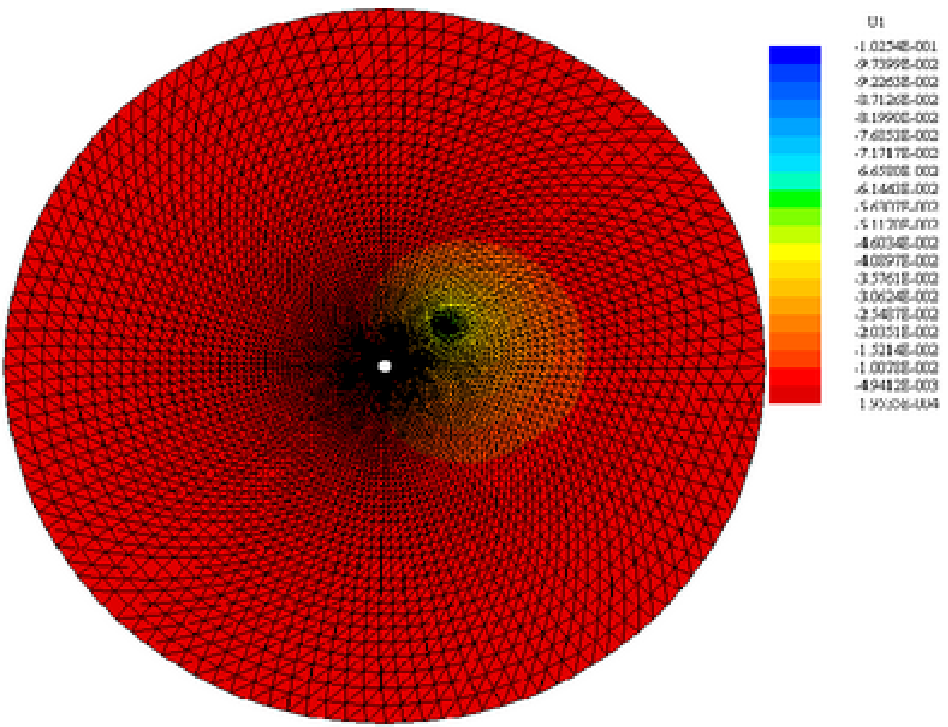}} }
\caption{Scalar gravitational field $\Phi$ at time $t=5\,M$ (n=1000) [left]
and at time $t=20\,M$ (n=4000) [right]. \label{afemshot1}}
\end{figure}

\begin{figure}[!htb]
\centerline{
\parbox{3.in}{\includegraphics*[width=2.5in]{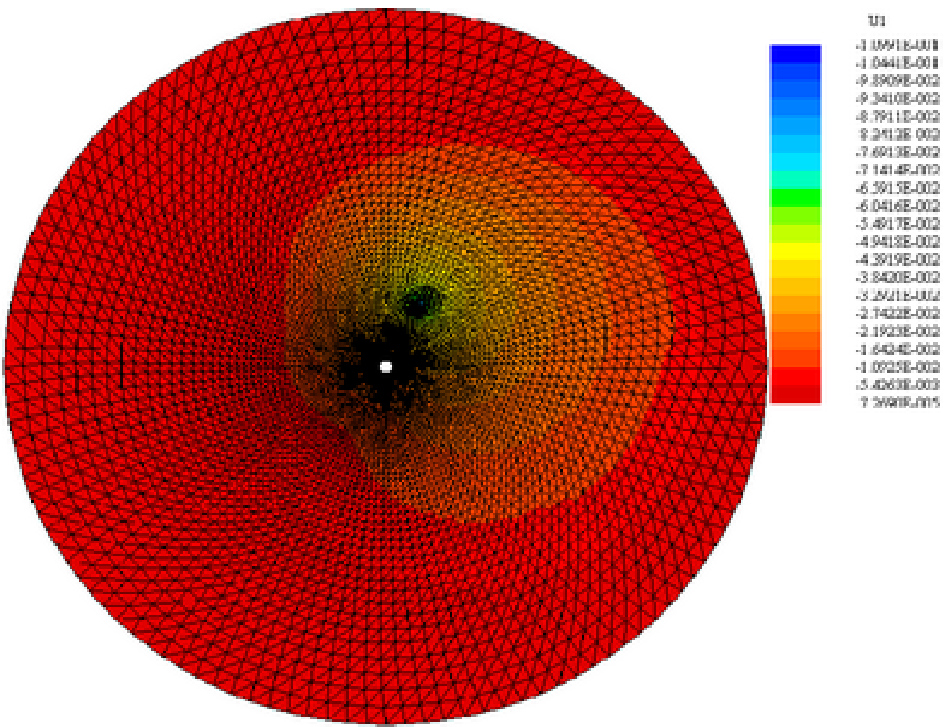}}
\parbox{3.in}{\includegraphics*[width=2.5in]{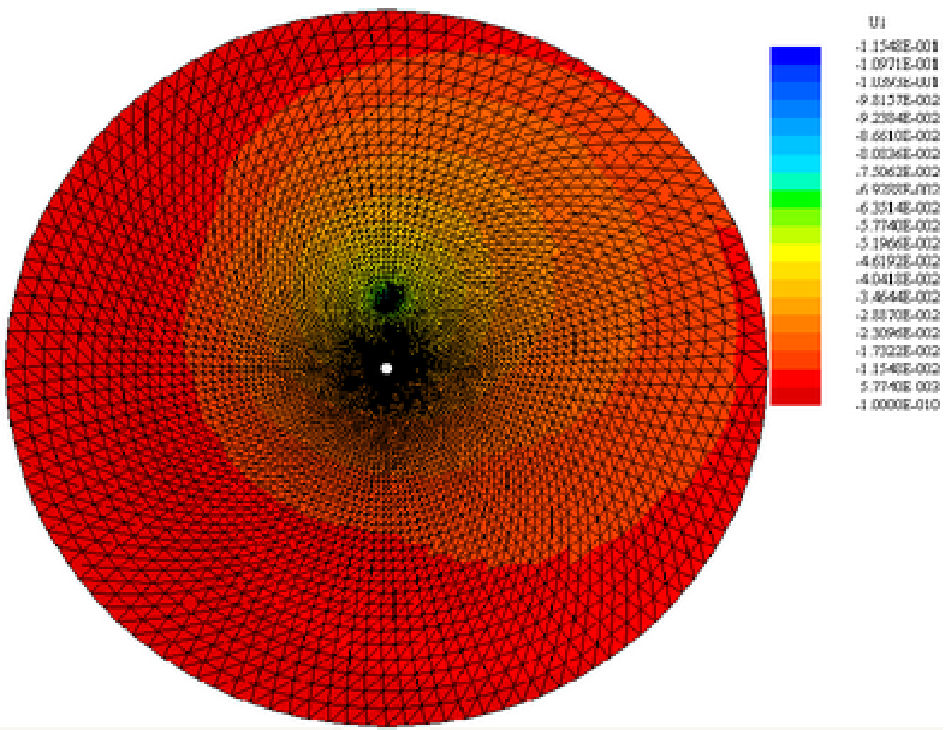}} }
\caption{Scalar gravitational field $\Phi$ at time $t=35\,M$ (n=7000) [left]
and at time $t=50\,M$ (n=10000) [right]. \label{afemshot2}}
\end{figure}

\begin{figure}[!htb]
\centerline{
\parbox{3.in}{\includegraphics*[width=2.5in]{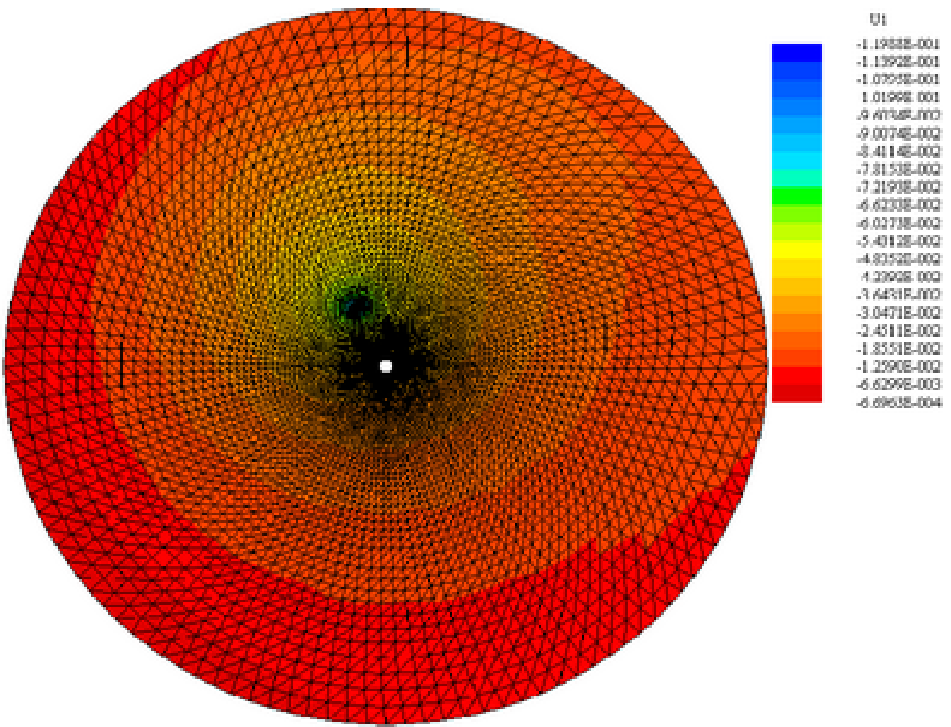}}
\parbox{3.in}{\includegraphics*[width=2.5in]{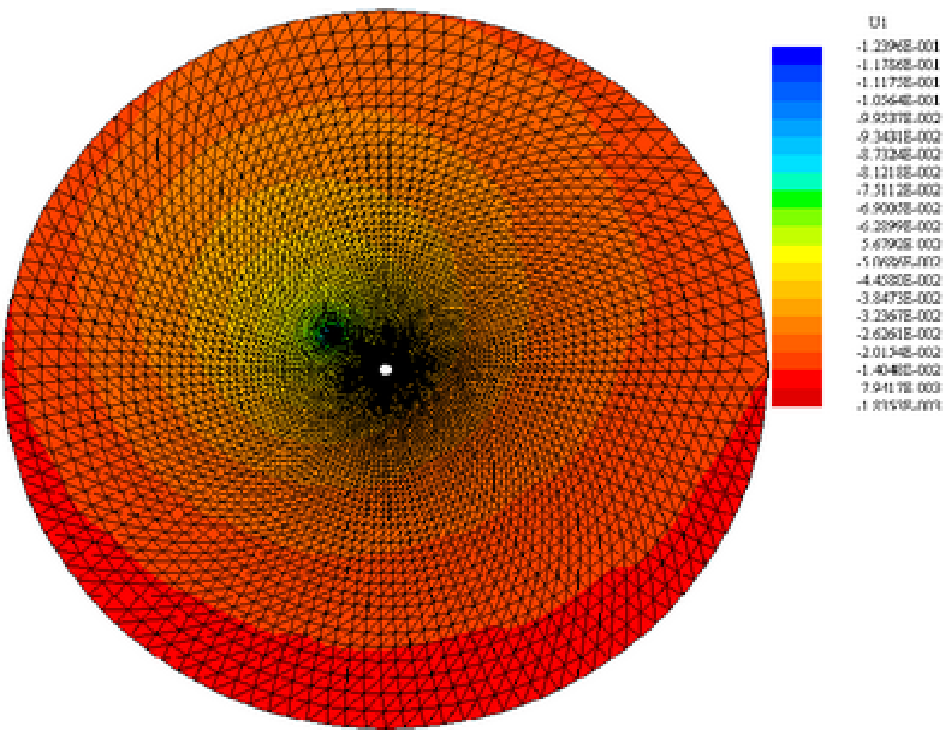}} }
\caption{Scalar gravitational field $\Phi$ at time $t=65\,M$ (n=13000) [left]
and at time $t=80\,M$ (n=16000) [right]. \label{afemshot3}}
\end{figure}

\begin{figure}[!htb]
\centerline{
\parbox{3.in}{\includegraphics*[width=2.5in]{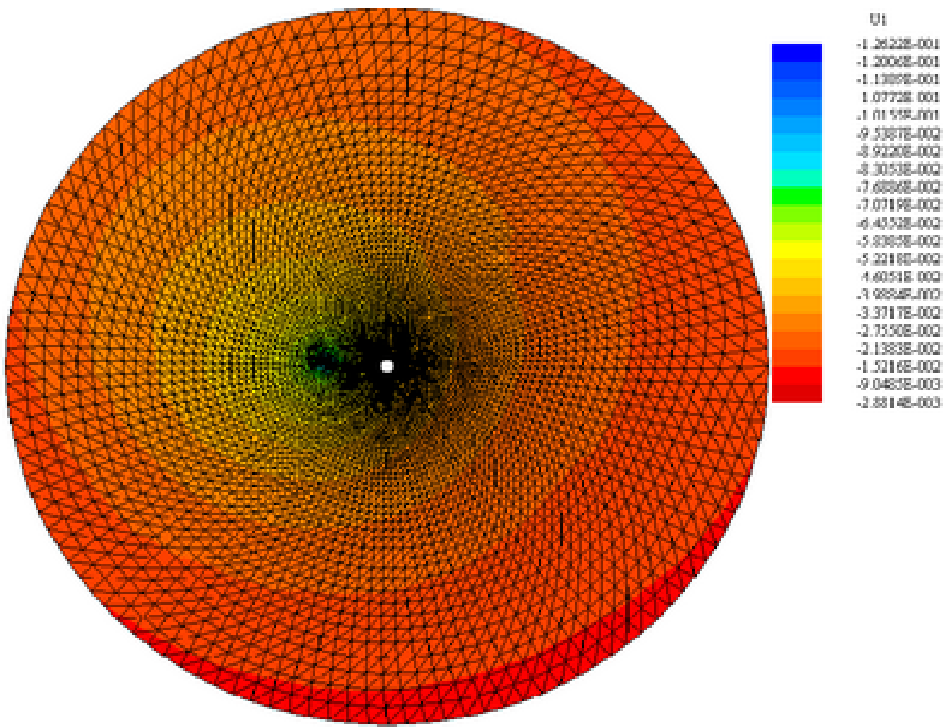}}
\parbox{3.in}{\includegraphics*[width=2.5in]{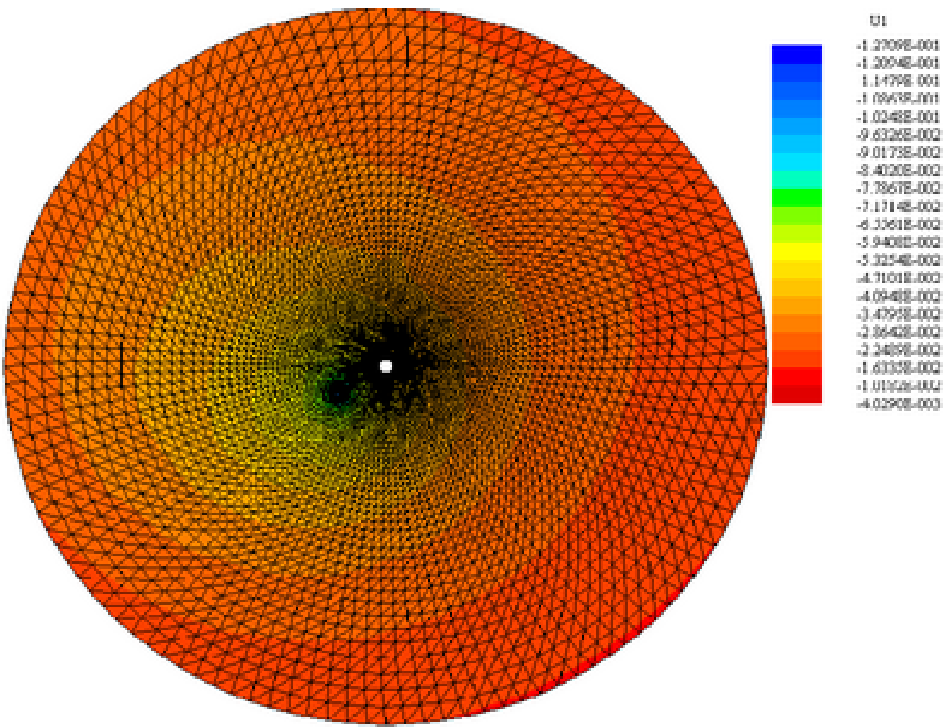}} }
\caption{Scalar gravitational field $\Phi$ at time $t=95\,M$ (n=19000) [left]
and at time $t=110\,M$ (n=22000) [right]. \label{afemshot4}}
\end{figure}

\begin{figure}[!htb]
\centerline{
\parbox{3.in}{\includegraphics*[width=2.5in]{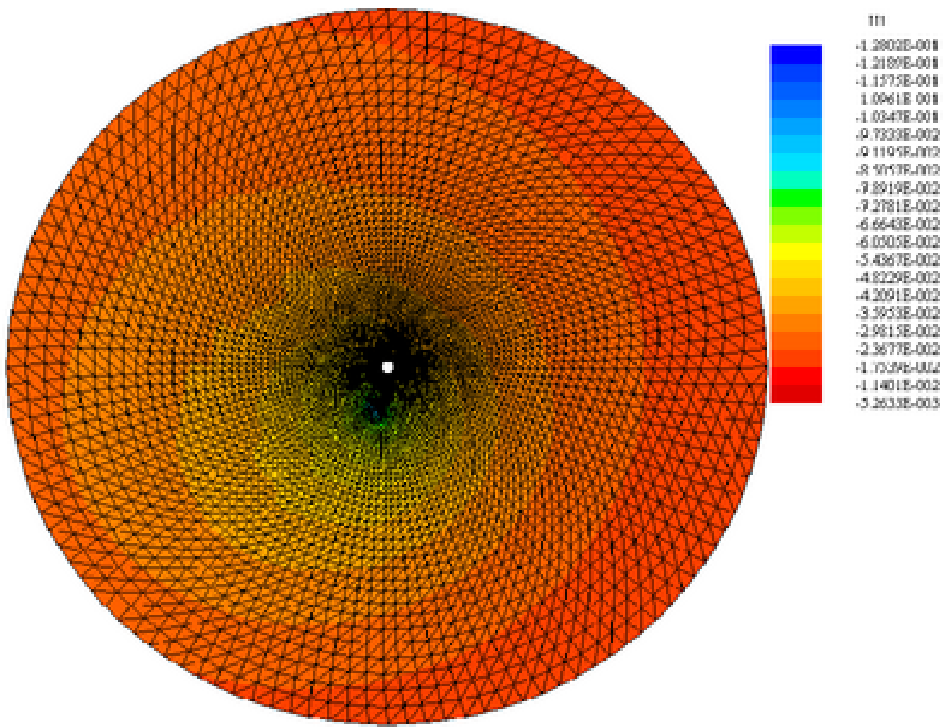}}
\parbox{3.in}{\includegraphics*[width=2.5in]{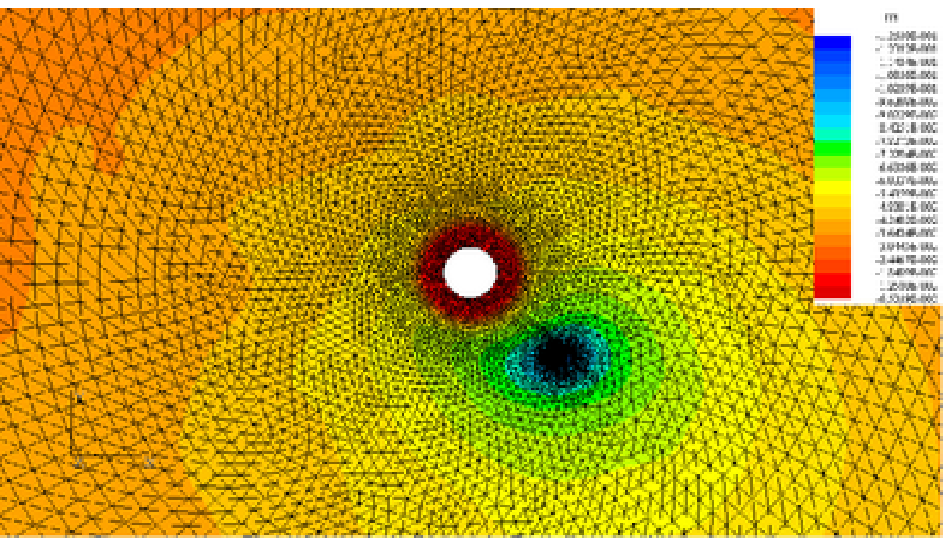}} }
\caption{Scalar gravitational field $\Phi$ at time $t=125\,M$ (n=25000) [left]
and at time $t=140\,M$ (n=30000) [right]. \label{afemshot5}}
\end{figure}

For the numerical simulations of the toy model with the AFEM we
use the same parameters as in subsection~\ref{without}, excepting
for the width of the Gaussian, which we have been able to reduce down to 
$\sigma=0.1\,M\,.$ And this has been done using a number of triangular
elements in the interval $11000-15000$ (the number of elements changes
with the evolution, depending on how the adaptivity is implemented),  
a number comparable to the one used
in the calculations without adaptivity.  That is, working with the AFEM we have 
been able to use a width $\sigma$ an order of magnitude smaller than with the 
classical FEM, using a comparable number of elements.
The time step is now $\Delta t =
0.005M\,,$ half the one used in the previous simulations.
The trajectory followed by the particle is drawn in Figure~\ref{trajectoryafem}.
We show the evolution of the scalar gravitational field in
Figures~\ref{afemshot1}-\ref{afemshot5}.  These are contour plots
as the ones we used previously. 
In order to show how we refine the mesh in the evolution we have superposed the 
mesh to the contour plots.  In this way we can see how the high-resolution 
part of the mesh moves with the particle.  
One can clearly see the difference
between this case and the previous case:  the particle inspirals 
much faster in this case, and it hardly completes more than one
orbit.  To understand this difference, it is important to remark the fact
that the particle sources are different since the Gaussian profiles have a
different width $\sigma$ and the fact that the sources couple non-linearly
with the scalar gravitational field.  As a consequence, the difference in
$\sigma$ leads to a different evolution.
We have checked that when take $\sigma=1M$ with the AFEM we recover the
type of trajectories we get in the case without adaptivity.

\begin{figure}[!htb]
\centerline{
\parbox{3.in}{\includegraphics*[width=2.5in]{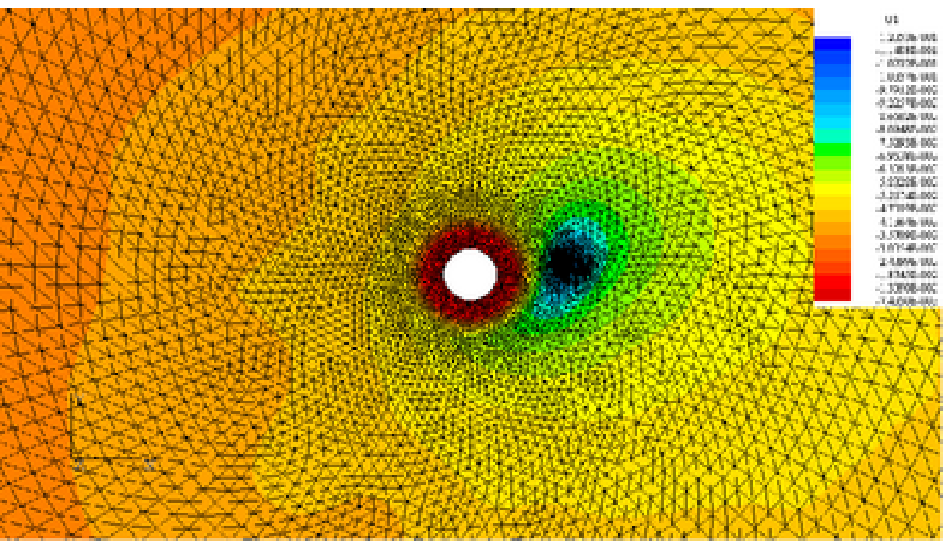}}
\parbox{3.in}{\includegraphics*[width=2.5in]{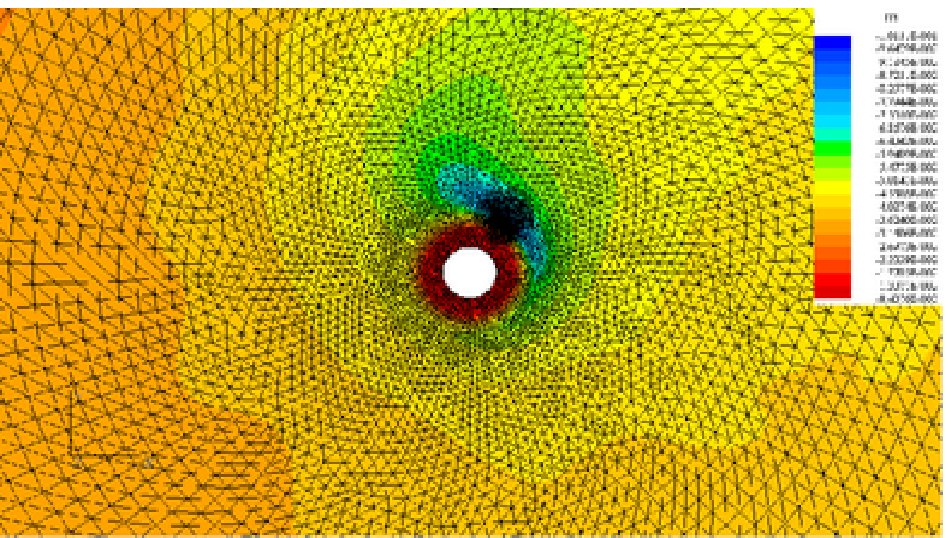}} }
\caption{Scalar gravitational field $\Phi$ at time $t=150\,M$ (n=25000) [left]
and at time $t=160\,M$ (n=32000) [right]. \label{afemshot6}}
\end{figure}

\begin{figure}[!htb]
\centerline{
{\includegraphics*[width=4.in]{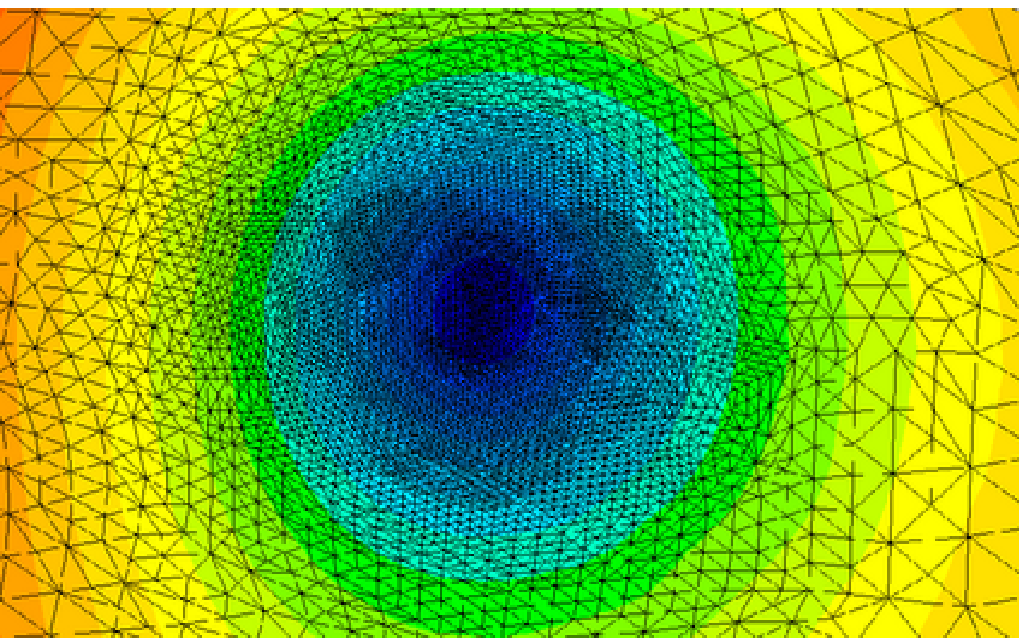}} }
\caption{Details of the refinement used around the particle's position.
\label{localref}}
\end{figure}

\begin{figure}[!htb]
\centerline{
{\includegraphics*[width=4.in]{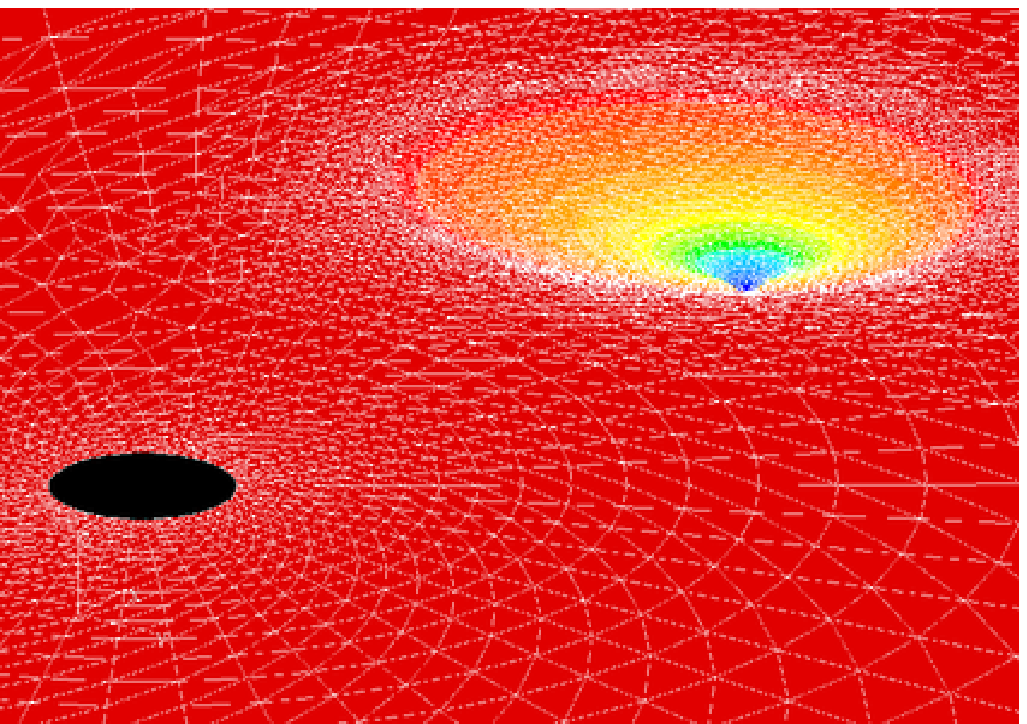}} }
\caption{A three-dimensional visualization of the mesh including
the excised region and the refinement around the particle. \label{3dplot}}
\end{figure}

A more detailed view of the structure of the refined region around the particle 
is shown in Figure~\ref{localref}, where one can see the different levels of
refinement, and how the resolution increases as one approaches the particle. 
A global perspective of the structure of the mesh, containing the
excised region around the singularity of the black-hole,
and the refined region around the particle can be seen in the
three-dimensional graph shown in Figure~\ref{3dplot}.

\begin{figure}[!htb]
\centerline{
{\includegraphics*[width=5in]{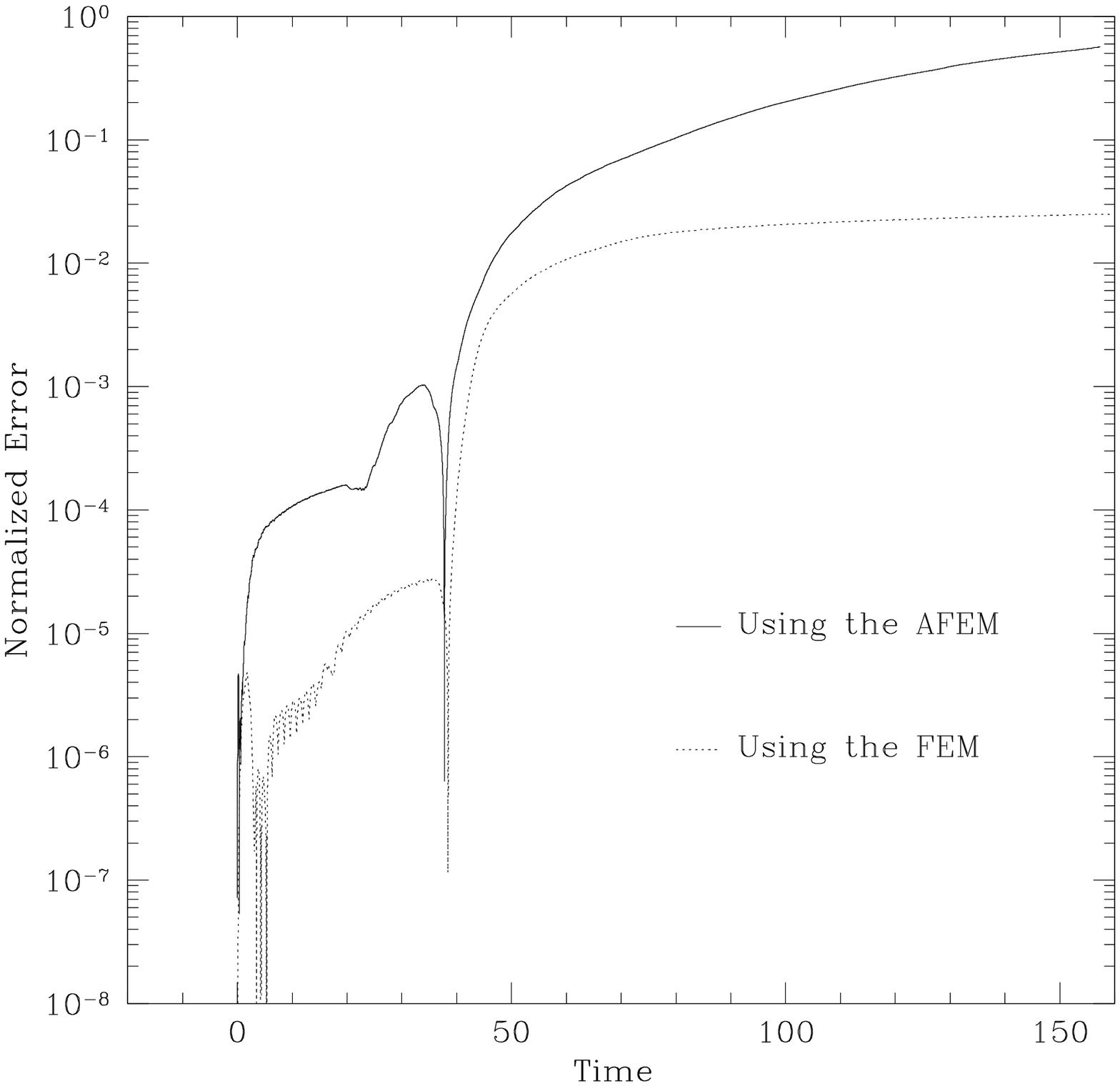}} }
\caption{Energy-Balance Test for the simulation that uses the AFEM. \label{etestwa}}
\end{figure}

We have also checked the energy-balance law~(\ref{claw}) for the simulations
with adaptivity.   We have plotted the result in Figure~\ref{etestwa}, where
again the horizontal axis denotes time and the vertical axis the value
of the left-hand side of (\ref{claw}) normalized with the mass $m$ of the
particle.   The behaviour of the error in the conservation law is very
similar to the one in the case without adaptivity, where the error in the 
energy-balance test grows after the boundary affects the particle.  
In this case, the error does not seem to stabilize as clearly as in the
previous case, and in average the error in the conservation law is bigger.
However, one has to take into account that the evolution in this case is
much faster since the particle has plunged very quickly.  Moreover, 
since the number of elements in both cases is comparable, but in the case
of the AFEM a considerable amount of them have been used to increase the
resolution around the particle, this means that the density of elements 
near the boundaries is less and
therefore if errors are propagated from there, they can affect more the
result than in the case without adaptivity.   This can be avoided by using
more computational power to push the boundaries far enough so that we can
neglect their effect during the significant part of the evolution.

\section{Remarks and Conclusions\label{conclusions}}
In this work we have studied the application of Finite Element techniques
to the time-domain numerical simulations of EMRBs.  More specifically, we have
shown how Finite Elements can help us in achieving the degree of adaptivity
that the computation of radiation reaction effects around the small body
requires.  
However, adaptivity is not the only advantage in using Finite Elements for
the study of EMRBs, some of which are obvious from the study of the toy
model we have considered in this work.  In short, the main additional advantages 
that Finite Elements provide are: 
(i) Versatility.  the FEM can be
applied to a wide range of problems: static, quasi-static, transient,
highly dynamical, linear and nonlinear, etc.  Moreover, the practical 
implementation of the FEM can be easily designed in a modular way, as
it shown by the existence of a number of multi-purpose Finite Element
packages.  This is a good property for the design of complex numerical 
codes, as the ones one may need for the description of astrophysical
EMRBs.
(ii) Many of the procedures that one uses in the framework of
the FEM have a solid theoretical foundation, in the sense that behind
them one often finds rigorous mathematical analysis (examples of this
can be found in the classical 
textbooks~\cite{StrFix:73,Zienkiewicz:77oc,Ciarlet_1978,Hughes:1987tj,Babuska:01bs}).
(iii) The ability of the FEM to manage problems 
with complex geometries.  For instance we can easily consider spherical boundaries
which is helpful in imposing boundary conditions and also in implementing
excision techniques for black hole singularities.  This is something that
may be worth to export to fully numerical relativistic calculations of
spacetimes containing black holes.  
(iv) In relation with the previous issue, the imposition of boundary 
conditions can be done in a somewhat natural way in a FEM framework.
The idea is that by adapting the mesh to the geometry of the problem under
study, boundary conditions which are also adapted to that geometry will
be incorporated in a natural way into the FEM discretization through the
{\em weak} formulation of the problem.  The outgoing boundary conditions that
we have used in our toy model are a good example, its implementation is
simpler that it would have been in a similar Finite Differences scheme.
(v) Another advantage of the FEM, that we have not explored in this paper,
is that it is to some extent natural to handle distributions, like
for instance the typical Dirac delta distributions that appear in 
physical systems containing particles (e.g., EMRBs).  The idea is that
we can use the exact properties of the distributions at the level
of the {\em weak} formulation of the equations, which in the case
of Dirac delta distributions leads automatically to a regularization
of the singularity associated with that kind of distributions.
We can apply this idea, for instance,  to the calculation of the 
general-relativistic perturbations of a non-rotating black hole, 
produced by the presence of a point-like object, via the 
Regge-Wheeler-Zerilli formalism.   Then, the type of FEM discretization 
we would get is similar to the one obtained by 
Price and Lousto~\cite{Lousto:1997wf,Martel:2001yf,Martel:2003jj},
where they also used an integral form of the equations in order to 
obtain a discretization.  In addition, the FEM procedure can be 
generalized to higher-dimensional problems in a straightforward way.
These ideas will be the subject of future investigations

In this paper we have illustrated the capabilities of the FEM and its 
suitability to simulate EMRBs, by applying it to a simplified model, 
a toy model that retains the basic ingredients and difficulties of 
general relativistic EMRBs.   
We have shown simulations of this system based on the classical 
FEM and simulations based on the AFEM.  An outcome of this work is the 
realization that in order to increase the resolution around the small body, 
treated as a particle, the introduction of adaptivity in the region of the
particle is necessary and in this sense the AFEM is a natural tool to use.  
The primary benefit of the AFEM is that it can provide an efficient, accurate 
and reliable computational analysis of very large continuum problems, 
for only a relatively small fraction of the cost associated with the 
non-adaptive FEM.   The accuracy of a finite element solution is directly 
dependent on the number of free parameters used to mathematically represent 
the problem, and how effectively those parameters, or mathematical degrees of 
freedom, are distributed over the problem space. Furthermore, the full 
computational cost associated with obtaining a finite element solution is 
related to both the number and the interconnectivity of the degrees of freedom 
used in the problem discretization. Consequently, the most efficient 
distribution of degrees of freedom for a problem is the one that provides a 
sufficiently accurate solution for the lowest number of free parameters.  
Currently, the only practical way to achieve this objective is by using adaptive 
solution strategies which are capable of cleverly evolving and improving an 
efficient distribution of degrees of freedom over the problem domain by 
establishing solution error distributions, and then adjusting or adding degrees 
of freedom to the discretization in order to correct them. 
By increasing the numbers of degrees of freedom only in the regions of 
higher error in the solution, it is possible to make the most significant 
improvement in the global accuracy of the finite element solution for the 
minimum additional computational cost. 
In this paper we employed the Hessian matrix of the solution to describe the 
solution error distribution, which can perfectly guide us to where and how we 
have to locally refine the mesh without any unnecessary pollution.  In this 
sense, it is important to emphasize that the rest of adaptive strategies 
available do not have this advantageous property.  Moreover, this comes without 
paying an extra price since the main advantages of the other adaptivity 
techniques are present in the AFEM that we have used.

The natural continuation of this work is the transfer of the technology used 
here to the general relativistic problem.  One possible way is to
consider the general framework of metric perturbations in the setup of the
3+1 decomposition, that is, to use 3D perturbative numerical relativity,
trying at the same time to profit from the experience gained in 3D full 
numerical relativity.  However, 3D calculations using the AFEM may be at 
present computationally too demanding and therefore, other avenues should 
be also explored.   Among them we can consider 1D calculations restricted 
to the case of a non-rotating MBH, by using the well-known Regge-Wheeler 
and Zerilli-Moncrief formalisms, or just by using a harmonic gauge, where
the computation of self-forces seems more natural.   This would be an
interesting benchmark to test further the FEM techniques in a 
general-relativistic context.  From here, one can study problems of 
more interest for gravitational wave physics related to LISA (involving
spinning MBHs) by considering 2D calculations using the curvature based 
formalism of Teukolsky for linear perturbations around Kerr black holes.  
One can go beyond by considering also perturbations of Kerr black holes 
using metric perturbations in a harmonic gauge, where the computation of
self-forces can be carried out by using techniques already present in
the literature.

\section{Acknowledgements}
CFS and PL acknowledge the support of the Center for Gravitational
Wave Physics funded by the National Science Foundation under Cooperative
Agreement PHY-0114375.
This work was partially supported by NSF grant PHY-0244788  to Penn
State University.


\end{document}